\documentclass[aps,amsmath,superscriptaddress,twocolumn,footinbib,prx]{revtex4-2}
\usepackage{graphicx}
\usepackage{dcolumn} 
\usepackage{bm}
\usepackage[utf8]{inputenc}
\usepackage{amsmath}
\usepackage{amsthm}
\usepackage{hyperref}
\hypersetup{colorlinks,allcolors=blue}
\usepackage{physics}
\usepackage{mymacros}
\usepackage{natbib}
\usepackage{amsfonts}
\usepackage{tikz,pgf}
\usepackage{tabularx,stackengine}
\usepackage{soul}

\usetikzlibrary{decorations.markings, svg.path, decorations.pathmorphing, shapes.misc}
\usetikzlibrary{chains}
\usetikzlibrary{calc}
\definecolor{internationalkleinblue}{rgb}{0.0, 0.18, 0.65}

\newcommand{\im}[1]{\mathrm{im}\,#1}
\newcommand{\Z}{\mathcal{Z}}
\newcommand{\ZZ}{\mathbb{Z}}
\newcommand{\ft}{\mathbb{F}_2}
\newcommand{\R}{\mathcal{R}}
\renewcommand{\D}{\mathcal{D}}
\renewcommand{\C}{\mathcal{C}}
\renewcommand{\AA}{\mathsf{A}}
\renewcommand{\MM}{\mathsf{M}}

\begin{document}
\tikzset{arr/.append style={
        decoration={markings,
            mark= at position {#1} with {\arrow{>}} ,
        },
        postaction={decorate}
    }
}
\tikzset{midarr/.append style={
        decoration={markings,
            mark= at position .5 with {\arrow{>}} ,
        },
        postaction={decorate}
    }
}

\title{Preparing topological states with finite depth simultaneous commuting gates}
\author{Yarden Sheffer}
\affiliation{Department of Condensed Matter Physics, Weizmann Institute of Science Rehovot 7610001, Israel}
\author{Erez Berg}
\affiliation{Department of Condensed Matter Physics, Weizmann Institute of Science Rehovot 7610001, Israel}
\author{Ady Stern}
\affiliation{Department of Condensed Matter Physics, Weizmann Institute of Science Rehovot 7610001, Israel}
\begin{abstract}
    We present protocols for preparing two-dimensional abelian and non-abelian topologically ordered states by employing finite depth unitary circuits composed of long-ranged, simultaneous, and mutually commuting two-qubit gates. Our protocols are motivated by recent proposals for circuits in trapped ion systems, which allow each qubit to participate in multiple gates simultaneously.
    Our circuits are shown to be optimal, in the sense that the number of two-qubit gates and ancilla qubits scales as $O(L^2)$, where $L$ is the linear size of the system. Examples include the ground states of the toric code, certain Kitaev quantum double models, and string net models. Going beyond two dimensions, we extend our scheme to more general Calderbank-Shor-Steane (CSS) codes.
    As an application, we present protocols for realizing the three-dimensional Haah's code and X-Cube fracton models.
\end{abstract}
\date{\today}
\maketitle

\section{Introduction}
Recent advances in quantum hardware enabled quantum simulation of topologically ordered quantum states \cite{satzinger2021realizing, PRXQuantum.4.020339, tantivasadakarn_shortest_2022, verresen_efficiently_2022, bravyi_adaptive_2022,andersen_observation_2022,iqbal2024non, xu2024non,minev2024realizing}. The interest in such states stems from their use as quantum error-correcting codes, as well as from their characteristic anyonic excitations, whose existence is hard to find and verify in solid-state systems.

A significant challenge of realizing such states on a quantum computer is that, by definition, the creation of a topologically ordered state from a trivial state requires a quantum unitary circuit of depth $O(L)$ where $L$ is the linear dimension of the system, when one uses a circuit that is composed only of (spatially) local gates \cite{chen2010local}. Even when giving up on spatial locality, and allowing any set of few-qubit gates (the so-called $k$-local circuit model), the circuit depth is at least $O(\log{L})$ (circuits satisfying this bound can be realized using a multiscale entanglement renormalization ansatz (MERA) construction \cite{PhysRevB.79.195123}). Such extensive depth increases the difficulty of realizing topologically ordered states in noisy intermediate-scale quantum devices. Therefore, it is of both theoretical and practical importance to ask whether it is possible to obtain \textit{finite-depth} circuits for creating topologically ordered states.

Recently, several theoretical works showed that topologically ordered states (including non-abelian ones) can be created in finite-depth circuits using a combination of unitary operations, measurements, and corrections based on the measurement outcomes (classical feedback) \cite{piroli_quantum_2021,lensky_graph_2023, tantivasadakarn_shortest_2022, verresen_efficiently_2022, bravyi_adaptive_2022}. The resulting protocols enabled quantum simulations of non-abelian states \cite{andersen_observation_2022,iqbal2024non}. Here we consider another path to obtaining finite-depth circuits for topologically ordered states, using simultaneous commuting gates. Certain cold-atoms computer architectures allow the application of an all-to-all unitary gate of the Ising interaction form \cite{PhysRevA.101.032330}
\begin{equation}
    U_{\rm ising} = \exp{i\sum_{i<j}a_{ij} Z_i Z_j+i\sum_i h_i Z_i}
\end{equation}
where $i,j$ sum over all qubits, $Z_i$ is the $Z$ operator on the $i$'th qubit, and $a_{ij},h_i$ are parameters controlled by the interaction strength. Here we consider the simpler case in which
a proper choice of $a_{ij}$ and $h_i$ gives
\begin{equation}
    \label{eq: cz prod}
    U_{\rm ising} = \prod_{\{ij\}} CZ_{i,j},
\end{equation}
with $CZ_{i,j} = e^{i\frac{\pi}{4}(1-Z_i)(1-Z_j)}$ being the controlled-$Z$ operations on qubits $i,j$, and $\{i,j\}$ is a set of pairs of qubits. Such an interaction can be combined with finite-depth operations to allow for every form of simultaneous commuting controlled unitary operations, where each unitary acts on an $O(1)$ number of qubits \cite{hoyer_quantum_2005}.

Interestingly, it was shown that measurements and non-local simultaneous commuting interactions are computationally equivalent. 
Specifically, Ref. \cite{browne_computational_2009} proved that a circuit of depth $D$ on $N$ qubits using local unitary operations, measurements, and feedback based on the measurement outcomes can be realized using local operations and unitaries of the form \eqref{eq: cz prod}, of depth and number of ancillary qubits being a constant multiplication of $D, N$.
Conversely, a circuit of depth $D$ on $N$ qubits using local operations and unitaries of the form \eqref{eq: cz prod} can be realized using local operations, measurements, and corrections using depth which is a constant multiplication of $D$ and $O(N^2)$ ancillas.

Furthermore, Ref. \cite{browne_computational_2009} gives a constructive method for converting finite-depth circuits with measurements and feedback to finite-depth circuits with simultaneous commuting gates (FDSC): every time an operator is measured we can instead save the result to an ancilla. Assuming for simplicity that a correction is a set of $Z$ operators applied on different qubits, we can, instead of applying the corrections classically based on the measurement outcomes, apply the corrections by quantum gates using the operator \eqref{eq: cz prod} applied on qubits to be corrected. Note that if the algorithm using measurements yields a deterministic result, tracing out the ancillas by the end of the algorithm will yield a pure state (as the resulting state on the computational qubits is independent of the values of the ancillas). In the context of the preparation of topologically ordered states, applying this method to the results of Refs. \cite{verresen_efficiently_2022, PRXQuantum.4.020339, bravyi_adaptive_2022}, one obtains (for two dimensional systems) finite-depth unitary circuits which require the application of $O(L^3)=O(N^{3/2})$ simultaneous two-body gates (where $L$ is the linear size of the system).

Here, we present FDSC protocols for the creation of abelian and non-abelian 2+1d topologically ordered (TO) states that require only $O(L^2)$ two-qubit gates and at most $O(L^2)$ ancillas (in fact, our protocols for abelian states require no ancillas at all). This is the optimal possible scaling for the number of gates with the system size.

To summarise our method: The TO ground state can be thought of as a superposition of all closed loops, where a line in a loop denotes that the qubit on the edge is in a particular state. The crucial step in the method is the division of the graph into a "spanning tree", i.e. a subgraph of the edges that contains no loops but goes through all vertices, and the complement of that tree. We then start with a product state in which qubits on the tree are all in the same state, and qubits on the complement are all in a different state. This initial state is a superposition of open strings, with all strings needing only one edge to be closed in order to form loops. We then ``close the loops" of the tree to demand a flux-free condition, obtaining the TO state. This is done by a controlled operation, in which each edge in the complement of the tree is assigned a single loop to be closed, and is controlled by the tree edges on that loop. We show that for certain TO states the controlled operation can be done using a finite depth circuit. The problem of obtaining an optimal number of ancillas and two-qubit gates then reduces to the problem of finding an optimal spanning tree. We give a construction for the tree that leads to an optimal scaling of the number of gates, being proportional to the number of qubits.

Let us review the outline of the paper. In section \ref{toric code} we begin by presenting simple protocols for generating the Greenberger-Horne-Zeilinger (GHZ) and toric code states using FDSC circuits. We then show in section \ref{optimal-preparation} that circuits for preparing the toric code (and more general topologically ordered states) can be optimized to contain only $O(L^2)$ two-qubit gates. The results of these two sections present, in a simple setting, the main ideas that will be used throughout the paper.

After setting the stage in the abelian example of the toric code, we move to more general examples. Sec. \ref{quantum double} generalizes the protocol to the non-abelian Kitaev quantum double model \cite{kitaev_fault-tolerant_2003}. It is shown that the ability to create the quantum double for a finite group $G$ is related to the question of multiplying a large number of group elements $g_1 \cdots g_n$ in FDSC. This naturally leads to the notion of a solvable group, presented before in a similar context \cite{PRXQuantum.4.020339,bravyi_adaptive_2022}. Next, we consider similar protocols for more general string-net models \cite{levin_string-net_2005}. In section \ref{string-nets}
we present the string-net model and discuss the general problem. In section \ref{abelian-string-nets} we present protocols for all abelian string nets, and in section \ref{string-nets-gauging} we present a procedure to obtain non-abelian string-nets from abelian ones using FDSC by a gauging procedure of an anyon-permuting symmetry with a solvable symmetry group $G$. 


Finally, we generalize our discussion of the toric code to Calderbank-Shor-Steane (CSS) codes in higher dimensions. In Sec. \ref{css-codes} we describe the generalization of our protocol to codes on arbitrary graphs. While the structure of the toric code, which allows us to obtain optimal protocols, does not exist in more general codes, we are able to give explicit protocols and explain how they can be optimized numerically. In Sec. \ref{fractons} we focus on two 3D fracton codes: the X-cube model \cite{vijay2016fracton} and Haah's code \cite{haah2011local}, and describe an FDSC protocol for realizing them. This provides an exciting possibility of near-term realization of fracton models in quantum devices. In Sec. \ref{outlook} we conclude our work and summarize possible future directions.

\section{GHZ and Toric Code States}
\label{toric code}
As the first examples, we describe how the GHZ and toric code states can be realized using a sequence of finite-depth simultaneous-commuting unitaries (FDSC). Without using FDSC gates, both states are known to require at least $O(\log N)$-depth circuits to be generated, even if one allows for $k$-local gates (a simple proof is given in Lemma 3.16 of \cite{nirkhe2022lower}, see also \cite{aharonov2018quantum}). 
Importantly, in order to achieve \textit{finite-depth} circuits using simultaneous gates, some qubits have to participate in an extensive number of gates. Otherwise, the unitary can be ``shrunk'' to a finite-depth circuit with no simultaneous gates acting on the same qubit.

\subsection{GHZ state}
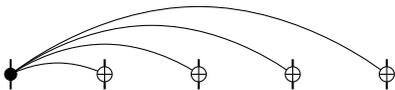
\begin{figure}
    \centering
    \begin{tikzpicture}
        \foreach \x in {0,1.25,...,5}{
            \draw[thick] (\x,-.2) -- (\x,.2);
        }
        \filldraw (0,0) circle (.08);
        \foreach \x in {1.25,2.5,...,5}{
            \draw plot [smooth,tension=1] coordinates {(0,0) (\x/2,\x*.2-.1) (\x,0)};
            \def\r{.1};
            \filldraw[fill=white] (\x,0) circle (\r);
            \draw (\x,-\r) -- (\x,\r);
            \draw (\x-\r,0) -- (\x+\r,0);
        };
    \end{tikzpicture}
    \caption{The circuit realising $U_{\rm GHZ}$}
    \label{fig:GHZ}
\end{figure}

We begin with the GHZ state:
\begin{equation}
    \ket{\rm GHZ} = \frac{1}{\sqrt{2}}(\ket{0\cdots 0} + \ket{1\cdots 1})
\end{equation}
on $N$ qubits. It can be similarly be defined as the stabilizer state of the operators $Z_1 Z_2, Z_2 Z_3,...,Z_{   N-1}Z_N$ and $X_1\cdots X_N$. An equivalent, but more useful to our protocol, set of stabilizers can be written as $Z_1 Z_2, Z_1 Z_3, ..., Z_1 Z_N, X_1\cdots X_N$. The state can then be created from the product state stabilized by the operators $X_1,Z_2,Z_3,...,Z_N$ using the operator
\begin{equation}
    U_{\rm GHZ} = \prod_{i=2}^N CX_{1,i},
\end{equation}
where $CX_{i,j}$ is the controlled-not operator, controlled by qubit $i$ and applied on qubit $j$:
\begin{equation}
    CX_{i,j} = e^{\frac{i \pi}{4}(1-Z_i)(1-X_j)}, 
\end{equation}
and we used the identities
\begin{equation}
    \begin{aligned}
    CX_{1,2} X_1 &= X_1 X_2 CX_{1,2}, \\
    CX_{1,2} Z_2 &= Z_1 Z_2 CX_{1,2}
    \end{aligned}
\end{equation}
(see Fig. \ref{fig:GHZ}). Since $\qty[CX_{1,i},CX_{1,j}]=0$ the $U_{\rm GHZ}$ is FDSC.

The physical interpretation of the above gate can be described as creating the ``information'' on a single qubit, and then entangling an extensive number of qubits to it.

\subsection{Toric code}
 The same idea can be used to create the toric code state. Here we present a simple but sub-optimal circuit, which will be improved in the next section. 
 
\begin{figure}
    \centering
    \begin{tikzpicture}[every node/.append style={font={\scriptsize}},scale=.8]
        \tikzstyle{thn}=[dotted,thick]
        \def\sz{4}
        \foreach \x in {0,...,\sz}{
            \draw[thn] (\x,-.3) -- (\x,\sz+.3);
            \draw[thn] (-.3,\x) -- (\sz+.3,\x);
            \draw[orange,very thick] (0,\x) -- (\sz,\x);
            }
        \node at (.5,\sz+.2) {$\ket{+}$};
        \node at (1.3,\sz-.5) {$\ket{0}$};
        \draw[orange,very thick] (0,0) -- (0,\sz);
    \end{tikzpicture}
    \caption{Initial product-state configuration of the protocol to generate the toric code. $\ket{+}$ states are defined on the qubits on a tree in the edge graph (in orange), and $\ket{0}$ states are defined on all other edges.}
    \label{fig:initial-tc}
\end{figure}
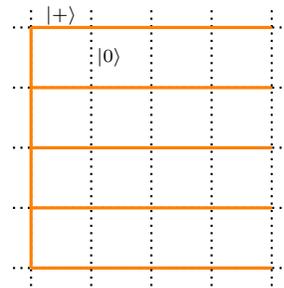

We begin with the product state on the edges of a square lattice, initialized such that all the horizontal edges and the edges at the leftmost vertical column are initialized in the $\ket{+}$ state, and all other qubits are initialized at $\ket{0}$ (see Fig. \ref{fig:initial-tc}). The edges initialized with $\ket{+}$ (orange in Fig. \ref{fig:initial-tc}) define a spanning tree of the square lattice. 

To obtain the toric code state, we need to ``close the loops'' of the tree, ensuring that each loop has trivial $\ZZ_2$ flux. For each loop $\ell$ closing the tree we define the loop operator $CL_\ell$ using a set of $CX$ operations as follows:
\begin{equation}
\label{eq:tc-cl}
CL_\ell=\vcenter{\hbox{
    \begin{tikzpicture}[every node/.style={font=\scriptsize}]
    \foreach \x in {1,...,4}{
        \draw[dashed] (\x,-.3) --   (\x,1.3);
        }
    \draw[thick,orange] (0,-.3) -- (0,1.3);
    \draw[thick,orange] (4.3,0) -- (0,0) -- (0,1) -- (4.3,1);
    \foreach \x in {0,...,3}{
        \filldraw (\x+.5,0) circle (.05);
        \filldraw (\x+.5,1) circle (.05);
        \draw[thin] (\x+.5,0) -- (4,.5);
        \draw[thin] (\x+.5,1) -- (4,.5);
    }
    \filldraw (0,.5) circle (.07);
    \draw (0,.5) -- (4,.5);
    \filldraw[fill=white] (4,.5) circle (.2);
    \draw (4-.2,.5) -- (4.2,.5);
    \draw (4,.3) -- (4,.7);
\end{tikzpicture}}}
\end{equation}
(that is, $CX$ operations controlled by the dotted qubits are applied on the $\oplus$ qubit). $CL_\ell$ guarantees that the flux inside the loop $\ell$ is trivial. Similar to $U_{\rm GHZ}$, $CL_\ell$ are FDSC. The FDSC unitary to create the toric code is therefore
\newcommand{\utc}{U_{\rm TC}}
\begin{equation}
\label{eq:tc-unitary}
    \utc = \prod_{\ell \in {\rm loops}} CL_\ell.
\end{equation}

This operation guarantees that all loops are flux-free, that is, the resulting state is stabilized by plaquette operators $B_p=\prod_{i\in p} Z_i$. We still need to ascertain that the state is stabilized by the vertex operators. One can see that, in terms of stabilizers,
\begin{equation}
    \utc \qty(
    \vcenter{\hbox{\begin{tikzpicture}[scale=.8]
        \draw[orange, very thick] (0,0) --node[fill=white] {\footnotesize $X$}  (-1,0);
        \draw[orange, very thick] (0,0) -- node[fill=white] {\footnotesize $X$}  (1,0);
        \draw[thin,dashed] (0,0) -- (0,1);
        \draw[dashed] (0,0) -- (0,-1);
    \end{tikzpicture}}}
    )\utc^\dagger=\qty(
    \vcenter{\hbox{\begin{tikzpicture}[scale=.8]
        \draw[orange,very thick] (0,0) --node[fill=white] {\footnotesize $X$}  (-1,0);
        \draw[orange, very thick] (0,0) -- node[fill=white] {\footnotesize $X$}  (1,0);
        \draw[orange, very thick] (0,0) -- node[fill=white] {\footnotesize $X$}  (0,-1);
        \draw[orange, very thick] (0,0) -- node[fill=white] {\footnotesize $X$}  (0,1);
    \end{tikzpicture}}}
    ),
\end{equation}
which ensures that the new state is stabilized by the toric code vertex operators $A_v=\prod_{i\in v} X_i$. 

This construction can be interpreted in terms of a potential function for the $\ZZ_2$ gauge field. Since the toric code is defined as a superposition of flux-free configurations of a $\ZZ_2$ gauge field $A_e$, we can define (at least locally) a $\ZZ_2$ ``potential'' $\Phi_v$ living on the vertices of the graph, such that 
\begin{equation}
\label{eq:phi_def}
    A_e=\Phi_{v_0}\Phi_{v_1}^{-1}
\end{equation} 
with $\Phi_{v_0,v_1}$ being the endpoints of the edge $e$. The above protocol can then be seen as first defining $\Phi_v$ on every vertex $v$ by choosing an arbitrary vertex $v_0$ and integrating over the path from $v_0$ to $v$ on the spanning tree
\begin{equation}
    \Phi_v = \prod_{e \in (\textrm{path from $v_0$ to $v$})} Z_e,
\end{equation}
then using the FDSC circuit to ensure that \eqref{eq:phi_def} is satisfied for all edges on the torus (not just those in the tree). This interpretation will prove useful in what follows. In particular, one notices that the toric-code ground state can be defined as an equal-weight superposition of all possibilities for $\Phi_v$. Since $\Phi_v$ can be defined using any spanning tree on the edge lattice, not necessarily that of Fig. \ref{fig:initial-tc}, any spanning tree gives rise to a similar protocol.

\section{Optimal number of gates for ground state preparation}
\label{optimal-preparation}
The finite-depth protocols we presented above require the use of $O(L^3)$ commuting two-qubit gates to obtain a topologically-ordered ground state on a lattice of length $L$. This raises the question of whether this scaling is the best possible for FDSC circuits for creating the toric code. Here we show that the number of gates can, in fact, be reduced significantly, by providing a protocol that requires only $O(L^2)$ two-qubit gates.

Let us stress some constraints on such construction. It is clear that $O(L^2)$ gates are optimal, as a smaller number would mean that some qubits are not entangled at all during the creation of the state. On the other hand, thinking of the ``light-cone'' of a qubit as the number of qubits it interacts with (assuming a single use of the interaction $U_{\rm ising}$), we see that at least some of the qubits must have a light-cone of size $L$, otherwise the expectation values of the toric-code loop operators could not be specified. Requiring only $O(L^2)$ two-qubit gates means that for any constant $K$ only $O(L)$ can have a light cone of size $>L/K$. One can say that the ``burden of entanglement'' has to be carried on by a small number of qubits.

We now present the protocol. In \ref{toric code} we showed that creating the toric code can be done by placing $\ket{+}$ states on any spanning in the edge graph, then using entangling gates to close the loops in the tree. Optimizing the number of two-qubit gates then amounts to choosing a spanning tree which minimizes the total length of the loops. 

\begin{figure}
    \centering
    \includegraphics[scale=.9]{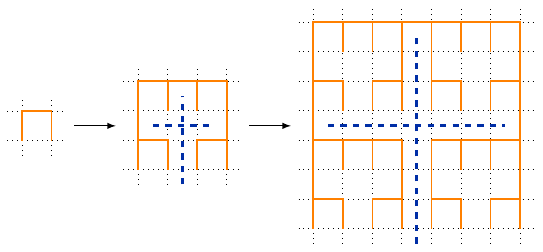}
    \caption{Construction of the self-similar trees. The $k$'th tree is constructed from 4 copies of the $k-1$ tree, attached on three edges. The blue crosses mark the edges whose loop closing requires additional operations besides the ones done on the sub-trees.}
    \label{fig: trees}
\end{figure}

We consider only lattices of size $L_k=2^k$. The spanning trees are defined recursively as in Fig. \ref{fig: trees}: The tree of size $L_k$ is constructed from four copies of the tree of size $L_{k-1}$, which are attached on the top, left, and right. The number $n_k$ of two-qubit gates required to close the loops on the $k$'th tree (excluding the ones going around the torus) is obtained by the sum of the number of existing edges in each loop. We notice that $n_k$ satisfies
\begin{equation}
\label{eq: trees-bound}
    n_{k+1}\le 4n_k + 17L_k^2
\end{equation}
This is because for each tree we must close the loops in the 4 sub-trees, as well as all the loops on the ``cross'' connecting them (marked in blue in Fig. \ref{fig: trees}). Each loop on the right and left part of the cross requires at most $4L_k$ gates to close, and each loop closing at the bottom part requires at most $8L_k$ gates to close. As a result of the bound \eqref{eq: trees-bound}, we have $n_k\le 9L_k$ so $n=O(L^2)$. 

Notice that the algorithm obtained from the equivalence with measurement-based quantum computation also gives a circuit with $O(L^3)$ gates. The protocol presented above therefore presents an asymptotic advantage in that sense. Also, we note that explicit calculation shows that the current algorithm outperforms the protocol presented in \ref{toric code} only above $L=8$. Nevertheless, one can obtain an optimal algorithm for a given system size by minimizing over the possible spanning trees numerically.

\section{Kitaev Quantum Double}
\label{quantum double}
The Kitaev quantum double $D(G)$ \cite{kitaev_fault-tolerant_2003} is the generalization of the toric code state to a general finite group $G$ (possibly non-abelian). In the following, we show that $D(G)$ can be generated by an FDSC circuit for any \textit{solvable} group $G$. We begin by defining the model and follow by showing how our procedure for the toric code can be generalized for an FDSC for any solvable $G$. This result mirrors the known result for measurement-based protocols \cite{bravyi_adaptive_2022, PRXQuantum.4.020339}.
\subsection{Definition of the model}
The quantum double model is defined on a general directed graph on the plane (here we pick the square lattice for simplicity), with a Hilbert space $\ket{g}$ on each edge corresponding to elements of a group $G$. The model is defined using commuting plaquette and vertex operators. The plaquette operators ensure that the ``$G$ flux'' on each plaquette vanishes:
\begin{equation}
    B_p\ket{
    \vcenter{\hbox{
    \begin{tikzpicture}[scale=.8]
    \begin{scope}[thick]
        \draw[midarr] (0,0) -- node[left] {\tiny $g_1$} (0,1);
        \draw[midarr] (0,1) -- node[above] {\tiny $g_2$} (1,1);
        \draw[midarr] (1,0) -- node[right] {\tiny $g_3$} (1,1);
        \draw[midarr] (0,0) -- node[below] {\tiny $g_4$} (1,0);
    \end{scope}
    \end{tikzpicture}
    } }}
        =
    \delta_{g_1 g_2 g_3^{-1} g_4^{-1}} \ket{
    \vcenter{\hbox{
    \begin{tikzpicture}[scale=.8]
    \begin{scope}[thick]
        \draw[midarr] (0,0) -- node[left] {\tiny $g_1$} (0,1);
        \draw[midarr] (0,1) -- node[above] {\tiny $g_2$} (1,1);
        \draw[midarr] (1,0) -- node[right] {\tiny $g_3$} (1,1);
        \draw[midarr] (0,0) -- node[below] {\tiny $g_4$} (1,0);
    \end{scope}
    \end{tikzpicture}
    }
    }}
\end{equation}
where $\d_g$ is 1 if $g$ is the identity element and $0$ otherwise. To define the vertex operators we write
\begin{equation}
L_v^g \ket{
\vcenter{\hbox{
\begin{tikzpicture}[scale=.9]
    \begin{scope}[thick,every node/.append style={font=\tiny}]
        \draw[midarr] (-1,0) -- node[above] { $g_1$} (0,0);
        \draw[midarr] (0,0) -- node[above] { $g_3$} (1,0);
        \draw[midarr] (0,0) --  (0,1) node[pos=.7,right] { $g_4$};
        \draw[midarr] (0,-1)  --  (0,0) node[pos=.3,right] { $g_2$};
    \end{scope}
\end{tikzpicture}
}}
}
=
\ket{
\vcenter{\hbox{
\begin{tikzpicture}[scale=.9]
    \begin{scope}[thick,every node/.append style={font=\tiny}]
        \draw[midarr] (-1,0) -- node[above] { $g_1g^{-1}$} (0,0);
        \draw[midarr] (0,0) -- node[above] { $gg_3$} (1,0);
        \draw[midarr] (0,0) --  (0,1) node[pos=.7,right] { $gg_4$};
        \draw[midarr] (0,-1)  --  (0,0) node[pos=.3,right] { $g_2g^{-1}$};
    \end{scope}
\end{tikzpicture}
}}
},
\end{equation}
so that the vertex operator is given by
\begin{equation}
    A_v =\frac{1}{\abs{G}} \sum_{g\in G} L_v^g.
\end{equation}
One can check that the vertex operator is indeed a projector and that vertex and plaquette operators commute. The Hamiltonian is then given by
\begin{equation}
    H=-\sum_v A_v -\sum_p B_p.
\end{equation}
The toric code, described above, corresponds to the case of $G=\mathbb{Z}_2$.
\subsection{Circuits for creating $D(G)$ ground states}
The procedure to create the $D(G)$ ground state can be seen as a generalization of the toric code protocol described in Sec. \ref{toric code}. One defines the states $\ket{0}_G\equiv\ket{e}$ where $e$ is the group unit, and $\ket{+}_G\equiv\frac{1}{\sqrt \abs{G}}\sum_g\ket{g}$. The relevant generalization to the $CX$ operator is 
\begin{equation}
    CX^G\ket{g_1,g_2}=\ket{g_1,g_1g_2}
\end{equation}
Similarly to the toric code case, states on the spanning tree defined in Fig. \ref{fig:initial-tc} are initialized in the $\ket{+}$ state, and states on the other qubits $\ket{0}$ (the procedure can immediately be converted to the graphs presented in Sec. \ref{optimal-preparation}). Again, this defines a $G$ potential function
\begin{equation}
    \Phi_v = \prod_{e \in(\textrm{path from $v_0$ to $v$})} g_e,
\end{equation}
where for non-abelian $G$ the product should be ordered, and $g$ has to take in account the orientation of the edge $g_e$. The loop unitaries are defined similarly to the above, using the $CX^G$ operators, and giving the generalization of \eqref{eq:tc-cl} as
\newcommand{\blk}[1]{\textcolor{black}{#1}}
\begin{equation}
\begin{aligned}
    CL_\ell^G&\ket{\vcenter{\hbox{
    \begin{tikzpicture}[scale=.9,
                        every node/.append style={font=\tiny}]
        \tikzset{plus/.style={orange,thick,midarr}}
        \def\ed{.35};
        \draw[plus] (0,0) -- node[below] {\blk{$g_{5}$}} ++(1,0);
        \draw[plus] (1,0) -- node[below] {\blk{$g_{6}$}} ++(1,0);
        \draw[plus] (2,0) -- node[below] {\blk{$g_{7}$}} ++(1,0);
        \draw[dashed] (-\ed,0) -- (0,0);
        \draw[dashed] (-\ed,1) -- (0,1);
        \draw[plus] (0,-\ed) -- node[left] {\blk{$g_{4}$}} (0,1+\ed);
        \foreach \x in {1,2}
            \draw[midarr,dashed] (\x,-\ed) --  (\x,1+\ed);
        \draw[midarr,dashed] (3,-\ed) -- node[right] {$h$} (3,1+\ed);
        \draw[plus] (0,1) -- node[above] {\blk{$g_3$}} (1,1);
        \draw[plus] (1,1) -- node[above] {\blk{$g_2$}} ++(1,0);
        \draw[plus] (2,1) -- node[above] {\blk{$g_1$}} ++ (1,0);
        \draw[orange,thick] (3,0) -- ++ (\ed,0);
        \draw[orange,thick] (3,1) -- ++ (\ed,0);
    \end{tikzpicture}
    }}}\\
    =&\ket{\vcenter{\hbox{
    \begin{tikzpicture}[scale=.9,
                        every node/.append style={font=\tiny}]
        \tikzset{plus/.style={orange,thick,midarr}}
        \def\ed{.35};
        \draw[plus] (0,0) -- node[below] {\blk{$g_{5}$}} ++(1,0);
        \draw[plus] (1,0) -- node[below] {\blk{$g_{6}$}} ++(1,0);
        \draw[plus] (2,0) -- node[below] {\blk{$g_{7}$}} ++(1,0);
        \draw[dashed] (-\ed,0) -- (0,0);
        \draw[dashed] (-\ed,1) -- (0,1);
        \draw[plus] (0,-\ed) -- node[left] {\blk{$g_{4}$}} (0,1+\ed);
        \foreach \x in {1,2}
            \draw[midarr,dashed] (\x,-\ed) --  (\x,1+\ed);
        \draw[plus] (3,0) -- node[right] {\blk{$hg_7^{-1}g_6^{-1}g_5^{-1}g_4g_3g_2g_1$}} (3,1);
        \draw[dashed] (3,0) -- (3,-\ed);
        \draw[dashed] (3,1) -- (3,1+\ed);
        \draw[plus] (0,1) -- node[above] {\blk{$g_3$}} (1,1);
        \draw[plus] (1,1) -- node[above] {\blk{$g_2$}} ++(1,0);
        \draw[plus] (2,1) -- node[above] {\blk{$g_1$}} ++ (1,0);
        \draw[orange,thick] (3,0) -- ++ (\ed,0);
        \draw[orange,thick] (3,1) -- ++ (\ed,0);
    \end{tikzpicture}
    }}},
    \end{aligned}
\end{equation}
such that \eqref{eq:tc-unitary} generalizes to
\begin{equation}
    U_{D(G)} = \prod_\ell CL_\ell^G.
\end{equation}
Since different $CL_\ell^G$ operators commute we have no issue in applying them simultaneously. The only issue in applying the procedure is that, naively, one has to apply the group multiplication $g_1\cdots g_n$ sequentially, without being able to take advantage of the simultaneous application of commuting gates, leading to $O(L)$-depth circuits. One can improve the protocol for a general group $G$ by multiplying the group elements with a ``divide and conquer'' algorithm, calculating every product $g_1\cdots g_n$ by dividing the calculation to $g_1\cdots g_{n/2}$ and $g_{n/2+1}\cdots g_n$, resulting in $O(\log L)$ depth. We wish, however, to obtain \textit{constant} depth circuits. In the following, we show that such a constant-depth protocol exists when $G$ is a solvable group.

\subsection{Constant depth multiplication protocol for solvable groups}
We begin with a definition: A group $G$ is solvable if there exists a sequence of groups ${e}=G_0\subseteq G_1 \subseteq ... \subseteq G_k=G$ such that $G_i$ is a normal subgroup of $G_{i+1}$ and the quotient group $H_i=G_i/G_{i-1}$ is abelian (a normal subgroup $N\subseteq G$ is a subgroup such that if $a\in N$ then $gag^{-1}\in N$ for any $g\in G$). The smallest $k$ satisfying this property is called the \textit{derived length} of $G$. To show a constant-depth protocol for multiplying the elements $g_1\cdots g_n$ can be realized for solvable groups, it is sufficient to show that if $N$ is a normal subgroup of $G$ such that $H=G/N$ is abelian, and multiplication in $N$ can be done in depth $D$, then multiplication in $G$ can be done in depth $D+c$ for $c$ constant (independent of the system size). The statement for a general solvable group then follows by induction over the derived length, yielding an $O(k)$-depth protocol. 

Before going to the details of the calculation, we comment on the picture behind it. A solvable group is "close" to a product of abelian groups $H_1\times\cdots\times H_k$, but with some nontrivial action between them, which makes the resulting group non-abelian. The method we present here relies on "isolating the non-abelian part" of the product, such the product can be written as a product of abelian terms, whose calculation requires some nontrivial functions of the group element. However, as each such function acts only on a small number of elements, they can be done in parallel using a pre-determined table of values.

Let us proceed with the proof. We let $\tau$ be the operation taking an element $g$ to its quotient in $H$, define a function $\psi:H\to G$ such that $\tau(\psi(h))=h$. Every $g\in G$ can then be written uniquely as $g=\psi(h)n$ for some $h\in H,n\in N$. By explicit calculation we have
\begin{equation}
\label{eq:solvable-product}
    \begin{aligned}
        g_1\cdots g_n &= \psi(h_1)n_1\cdots \psi(h_n)n_n \\
                      &= \psi(h_1\cdots h_n)\chi(h_1,h_2\cdots h_n)\varphi_{h_2\cdots h_n}(n_1)\\
                      &\times \chi(h_2,h_3\cdots h_n)\varphi_{h_3\cdots h_n}(n_2)\\
                      &\times\cdots\\
                      &\times \chi(h_{n-1},h_n) \varphi_{h_n}(n_{n-1})n_n.
    \end{aligned}
\end{equation}
where we defined the functions
\begin{align}
    \chi(h_1, h_2) &= \psi(h_1h_2)^{-1} \psi(h_1)\psi(h_2),\\
    \varphi_{h}(n) &= \psi^{-1}(h)n\psi(h)\in N
\end{align}
This expansion is valid for any normal subgroup $N\subseteq G$, independently of whether the quotient $H$ is abelian. We also have
\begin{equation}
    \tau(\chi(h_1, h_2))= e_H
\end{equation}
(the trivial element in $H$) so $\chi(h_1, h_2)\in N$. Note that $\chi$ is generally non-trivial as $\psi$ is not a group homomorphism. Besides the $\psi$ terms at the beginning of \eqref{eq:solvable-product}, all terms are elements of $N$, and can therefore be multiplied in constant depth by assumption. We now use the condition that $H$ is abelian so that the arguments for $\chi$ and $\varphi$ can be calculated similarly using FDSC operators. We can, therefore, calculate first the arguments for the functions $\chi,\varphi$, then calculate the product in constant depth. For a general solvable group, the multiplication algorithm is obtained recursively.

We note that the algorithm requires storing the results of the products $h_ih_{i+1}\cdots h_n$, as well as the outputs of the functions $\chi,\varphi$ in additional ancillas, requiring $O(n)$ ancillas for storage where $n$ is the number of elements to be multiplied. For the tree structure presented in Sec. \ref{optimal-preparation} the total number of required ancillas can be bounded by a constant times the total length of the loops, giving $O(L^2)$ total ancillas.

\subsubsection*{Example: multiplication in the group $D_n$}
As an example of the above procedure, we consider the simplest class of solvable non-abelian groups: the groups $D_n$ of symmetries of a regular $n$-sided polygon on the plane. The group elements are rotations and reflections, totaling $2n$ elements. The group $D_n$ has is a normal abelian subgroup, the group $N={r^k}$ of pure rotations. The coset group $H=D_n/N$ consists of two elements: the trivial element and a reflection $m$. The function $\psi$ above sends $m$ to one of the reflections on the plane. Since $\psi(m)r\psi(m)^{-1}=r^{-1}$ we obtain
\begin{equation}
    \begin{aligned}
        \chi(h_1,h_2) &= e,\\
        \phi_h(r^k)&=r^{(-1)^h}
    \end{aligned}
\end{equation}
where $(-1)^h$ is $-1$ if $h=m$ and 1 otherwise. Since $\chi$ is trivial we will use $m$ to denote $\psi(m)$ below. Defining $g_i=m^{p_i} r^{k_i}$, we see that \eqref{eq:solvable-product} reduces to
\begin{equation}
    \label{eq:Dn product}
    g_1\cdots g_m = m^{\sum_i p_i} \,r^{(-1)^{(\sum_{i\ge 2}p_i)}k_1} r^{(-1)^{(\sum_{i\ge 3}p_i)}k_2}\cdots r^{k_m},
\end{equation}
where sums over $p_i$ are taken mod 2. The formula above has a simple interpretation: the total product has a reflection if there is an odd number of reflections in its elements, and each rotation in the product elements contributes itself or its inverse depending on the parity of the number of reflections after it. The imprints of the non-abelian nature of the group are in the fact that the power of each rotation $r$ depends on the number of reflections to its right. 

Using \eqref{eq:Dn product} to calculate the product using FDSC gates can be done as follows: the sums $\sum_{i\ge j}p_i$ should be calculated and stored in ancillas. Since the elements are in $\ZZ_2$ the entire product can be calculated using a set of $CX$ gates. We then apply a set of parallel two-qudit gates to calculate the elements $r^{(-1)^{\sum_{i\ge j}p_i}k_{j-1}}$. Finally, all these elements are multiplied using a set of $\ZZ_n$ $CX$ gates.

\section{Constructing general anyon theories via string-net models}
\label{string-nets}
We wish to generalize our construction to a broader class of topologically ordered phases. Our goal is to show that the ground states of all abelian (doubled) topologically ordered states can be constructed using FDSC gates and that non-abelian states can be obtained explicitly from abelian ones by gauging an anyonic symmetry \cite{teo_theory_2015,barkeshli_symmetry_2019} (also referred to as equivariantization in the mathematical literature \cite{etingof_weakly_2009}). To proceed, we use the general string-net model \cite{levin_string-net_2005}, which can be constructed for any anyon theory $\C$. In this section we recall some general definitions for anyon theories and define the string-net model, this is covered below in subsections \ref{anyon-theory-definitions} and \ref{string-net-defs} and can be skipped by readers familiar with string-net models. We then show how the general loop construction we used in the previous cases can be applied to the string-net model. This construction requires $O(L)$-depth circuits in the general case, but can be improved for specific cases: We show how to construct FDSC circuits for all abelian doubled theories (section \ref{abelian-string-nets}). Non-abelian theories can be obtained from an abelian (or non-abelian) one by gauging a solvable group symmetry using an FDSC circuit (section \ref{string-nets-gauging}).

\subsection{Definitions}
\label{anyon-theory-definitions}
An anyon theory (or a braided fusion category) $\C$ is defined as a collection of anyons and their algebraic relations. To specify $\C$, we begin with a collection of anyon labels $a,b,...$. Two anyons $a,b$ can be fused, resulting in an anyon $c$, and the possible fusion outcomes are specified using the matrices $N_{ab}^c$ which are 1 if $a,b$ can be fused to obtain $c$ and 0 otherwise (we neglect the possibility of $N_{ab}^c>1$ for simplicity). The anyon theory contains a ``trivial'' (or vacuum) anyon $1$, which fuses trivially with other anyons, and each anyon $i$ has a unique inverse $\bar{i}$ such that $N_{i\bar{i}}^1=1$ (note that in a general anyon theory the fusion of $i,\bar{i}$ can also have additional outcomes $N_{i\bar{i}}^j\neq 0$). The $N$ matrices satisfy symmetry conditions related to the possibility of rearranging the edges around the vertex:
\begin{equation}
    N_{ij}^k = N_{ji}^k = N_{\bar{i}\bar{j}}^{\bar{k}}=N_{i\bar{k}}^j.
\end{equation}

Different diagrams describing anyon fusion are related by the so-called $F$ moves, defined by
\begin{equation}
\label{eq:f-move}
\vcenter{\hbox{
    \begin{tikzpicture}[scale=.6]
        \begin{scope}[thick,every node/.append style={font=\scriptsize}]
            \draw[midarr] (-1,1) -- node[above=.1] {$i$} (0,0);
            \draw[midarr] (-1,-1) -- node[below=.1] {$j$} (0,0);
            \draw[midarr] (1,0) -- node[above] {$m$} (0,0);
            \draw[midarr] (2,-1) -- node[below=.1] {$k$} (1,0);
            \draw[midarr] (2,1) -- node[above=.1] {$l$} (1,0);
        \end{scope}
    \end{tikzpicture}}}
    =\sum_n F^{ijm}_{kln} \qty(\vcenter{\hbox{
    \begin{tikzpicture}[scale=.5]
        \begin{scope}[thick,every node/.append style={font=\scriptsize}]
            \draw[midarr] (-1,2) -- node[above=.1] {$i$} (0,1);
            \draw[midarr] (-1,-1) -- node[below=.1] {$j$} (0,0);
            \draw[midarr] (0,0) -- node[right] {$n$} (0,1);
            \draw[midarr] (1,-1) -- node[below=.1] {$k$} (0,0);
            \draw[midarr] (1,2) -- node[above=.1] {$l$} (0,1);
        \end{scope}
    \end{tikzpicture}}}
    ),
\end{equation}
where we use the notation of \cite{levin_string-net_2005}. By convention, we set $F_{kln}^{ijm}=0$ when any of the vertices in \eqref{eq:f-move} does not satisfy the fusion rule. We will further be interested in the braiding relations of the theory. They are given by the ``$R$ matrices'' $R_{ab}^c$ which give the phase obtained from the anyons $a,b$ moving around each other, given that they fuse to $c$, that is
\begin{equation}
    \vcenter{\hbox{
    \begin{tikzpicture}[scale=.7]
        \begin{scope}[thick,every node/.append style={,font=\scriptsize}]
            \draw[arr={.7}] (-1,1) .. controls (1,.3) .. (0,0);
            \filldraw[white] (0,.6) circle (.2cm);
            \draw[arr={.7}] (1,1) .. controls (-1,.3) .. (0,0);
            \node at (-1,.4) {$a$};
            \node at (1,.4) {$b$};
            \draw[arr={.5}] (0,0) -- node[right] {$c$} (0,-1);
        \end{scope}
    \end{tikzpicture}
    }}
    =
    R_{ab}^c
    \vcenter{\hbox{
\begin{tikzpicture}[scale=.7,every path/.append style={thick},
every node/.append style={font=\scriptsize}]
        \draw[arr={.6}] (1,1) -- node [right] {$a$} (0,0);
        \draw[arr={.6}] (-1,1) -- node[left] {$b$} (0,0);
        \draw[arr={.5}] (0,0) -- node [right] {$c$} (0,-1);
\end{tikzpicture}
    }}.
\end{equation}
For a given set of $F$-symbols, the possible sets of compatible $R$-symbols are obtained by solving the "hexagon equations", namely
\begin{equation}
\begin{aligned}
    R_{ca}^e F^{\bar{a}\bar{c}e}_{d\bar{b}g}R_{cb}^g&=\sum_f F^{\bar{a}\bar{c}e}_{d\bar{b}f} R_{fc}^d F^{\bar{b}\bar{a}f}_{d\bar{c}g}, \\
    (R_{ca}^e)^{-1} F^{\bar{a}\bar{c}e}_{d\bar{b}g}(R_{cb}^g)^{-1}&=\sum_f F^{\bar{a}\bar{c}e}_{d\bar{b}f} (R_{fc}^d)^{-1} F^{\bar{b}\bar{a}f}_{d\bar{c}g}.
\end{aligned}
\end{equation}
Geometrically, these equations mean that anyon lines can be moved around such that all topologically equivalent braidings give the same result.
\subsection{String-net models}
\label{string-net-defs}
We now define the Levin-Wen string-net model \cite{levin_string-net_2005} and discuss some of its properties. Readers who are familiar with the model can freely skip this section. 

As input data to the string-net model, one gives an anyon theory $\C$ with its data in the form of fusion rules and $F$ symbols (note that the $R$-symbols are not specified as input data for the models). The model is defined on the honeycomb lattice (or similarly on any trivalent lattice), with some orientation defined for each edge. The Hilbert space consists of an anyon label on each edge. The Hamiltonian is a sum of commuting vertex and plaquette terms
\begin{equation}
\label{eq:h-string-net}
    H=-\sum_p B_p -\sum_v A_v.
\end{equation}
The vertex term projects on states satisfying the requirement that the edges around the vertex obey the fusion rules:
\begin{equation}
    A_v = \ket{\vcenter{\hbox{
        \begin{tikzpicture}[scale=.4,every path/.append style={thick},
every node/.append style={font=\scriptsize}]
        \draw[arr={.6}] (-1,1) -- node[left] {$i$} (0,0);
        \draw[arr={.6}] (1,1) -- node [right] {$j$} (0,0);
        \draw[arr={.5}] (0,0) -- node [right] {$k$} (0,-1);
    \end{tikzpicture}
    }}}=N_{ij}^k
    \ket{\vcenter{\hbox{
        \begin{tikzpicture}[scale=.4,every path/.append style={thick},
every node/.append style={font=\scriptsize}]
        \draw[arr={.6}] (-1,1) -- node[left] {$i$} (0,0);
        \draw[arr={.6}] (1,1) -- node [right] {$j$} (0,0);
        \draw[arr={.5}] (0,0) -- node [right] {$k$} (0,-1);
    \end{tikzpicture}
    }}}
\end{equation}
where an anyon label is replaced by its inverse if the arrow direction is opposite to the diagram above. 

To define plaquette operators, it is useful to begin by defining a general string operator which inserts a string $s$. To do so, it is useful to think of the model in terms of the ``fat lattice'': we allow strings that are not attached directly to the edges, but rather float near the edges, and are then combined to them using the $F$ moves \eqref{eq:f-move}. The string operator associated with a string $s$ is then given by
\begin{equation}
\label{eq: string-op}
    \qty(W_s)_{i_1 i_2 ...i_N}^{i_1' i_2' ... i_N'}(e_1...e_N) =\prod_{k=1}^N F_{\bar{s} i_{k-1}' \bar{i}_k'}^{e_k \bar{i}_k i_{k-1}}
\end{equation}
(see Fig. \ref{fig: string-net-creation}a for the relevant notation). That is, the string changes the anyon labels $i_1,...,i_n$ on the path of the string to the labels $i_1',...,i_n'$ obtained by fusing the initial labels with $s$. The result generally depends on the labels $e_1,\dots,e_N$ of the edges pointing out of vertices through which the string passes (here we focus on operators that fuse strings, but this is not the most general string operator of the model \cite{levin_string-net_2005}). String operators on the same path satisfy the fusion algebra
\begin{equation}
    W_i W_j =\sum_k N_{ij}^k W_k.
\end{equation}
To define the plaquette operators we start with the operators adding a string around the plaquette
\begin{equation}
    B^s_p \ket{\vcenter{\hbox{
    \begin{tikzpicture}[scale=.5]
        \newdimen\r
    \r=1cm
    \draw[thick] (330:\r) foreach \x in {30,90,...,330} { -- (\x:\r) };
    \foreach \x in {30,90,...,330}{
        \draw[thick] (\x:\r) -- (\x:1.5*\r);
    }
    \end{tikzpicture}
    }}}  =
    \ket{\vcenter{\hbox{
    \begin{tikzpicture}[scale=.5]
        \newdimen\r
    \r=1cm
    \draw[thick] (330:\r) foreach \x in {30,90,...,330} { -- (\x:\r) };
    \foreach \x in {30,90,...,330}{
        \draw[thick] (\x:\r) -- (\x:1.5*\r);
    }
    \draw[very thick,dashed,violet] (0,0) circle (.7);
    \node[font={\scriptsize},violet] at (.4,0) {$s$};
    \end{tikzpicture}
    }}}.
\end{equation}
The plaquette operator is then the sum
\begin{equation}
    B_p = \sum_s \frac{d_s}{\mathcal{D}^2}B_p^s.
\end{equation}
where $d_s$ is the quantum dimension of $s$, defined as the largest eigenvalue of the matrix $N_{s*}^*$, and $\DD=\sqrt{\sum_s d_s^2}$. This choice ensures that $B_p$ is a projector and that on the ground state of $H$ the string operators can be deformed, that is
\begin{equation}
\label{eq:topo-string}
    B_p\ket{\vcenter{\hbox{
    \begin{tikzpicture}[scale=.5]
    \newdimen\r
    \r=1cm
    \draw[thick] (330:\r) foreach \x in {30,90,...,330} { -- (\x:\r) };
    \foreach \x in {30,90,...,330}{
        \draw[thick] (\x:\r) -- (\x:1.5*\r);
    }
    \draw[very thick,dashed,violet] (280:1.5*\r) .. controls (270:1*\r) and (330:1*\r) ..
    (0:.7*\r) .. controls (30:1*\r) and (90:1*\r) .. (80:1.5*\r);
    \node[font={\scriptsize},violet] at (.4,0) {$s$};
    \end{tikzpicture}
    }}}=
    B_p\ket{\vcenter{\hbox{
    \begin{tikzpicture}[scale=.5]
    \newdimen\r
    \r=1cm
    \draw[thick] (330:\r) foreach \x in {30,90,...,330} { -- (\x:\r) };
    \foreach \x in {30,90,...,330}{
        \draw[thick] (\x:\r) -- (\x:1.5*\r);
    }
    \draw[very thick,dashed,violet] (260:1.5*\r) .. controls (270:1*\r) and (210:1*\r) ..
    (180:.7*\r) .. controls (180-30:1*\r) and (90:1*\r) .. (100:1.5*\r);
    \node[font={\scriptsize},violet] at (-.4,0) {$s$};
    \end{tikzpicture}
    }}}.
\end{equation}
This shows that the ground state of $H$ is topological. It is described by the ``Drienfeld double'' anyon theory $\Z(\C)$ obtained from $\C$ \cite{levin_string-net_2005,kassel2012quantum,teo_theory_2015}. For example, in the case where the strings correspond to the irreducible representations of a finite group $G$, the ground state is the same as the quantum double model described above $\Z({\rm Rep}(G))=D(G)$. In the case where $\C$ is a modular anyon theory (that is, no string in $\C$ braids trivially), the ground state is described by $\Z(\C)=\C\times\bar{\C}$ where $\bar{\C}$ is the time-reversal partner of $\C$.

For our purposes, the identity \eqref{eq:topo-string} is important as it implies that the ground state can be constructed by ensuring that the state is stabilized by long loops, that is
\begin{equation}
\begin{aligned}
    &\sum_{\{s_i\}}\frac{d_1\cdots d_n}{\mathcal{D}^{2n}}
    \vcenter{\hbox{
    \begin{tikzpicture}[scale=.7]
        \foreach \x in {0,...,3}{
        \begin{scope}[shift={({\x*sqrt(3)},0)}]
            \draw[thick] (330:1) foreach \t in {30,90,...,330} { -- (\t:1) };
        \end{scope}
        }
        \filldraw[white] ({sqrt(3)*2-.5},-1.1) rectangle ++ (1,2.2);
        \node at ({sqrt(3)*2},0) {$\cdots$};
        \foreach \x in {0,1,3}
            \draw[very thick,violet,dashed] ({\x*sqrt(3)},0) circle (.7);
        \node[violet,font={\scriptsize}] at (.4,0) {$s_1$};
        \node[violet,font={\scriptsize}] at ({sqrt(3)+.4},0) {$s_2$};
        \node[violet,font={\scriptsize}] at ({3*sqrt(3)+.4},0) {$s_n$};
    \end{tikzpicture}
    }}\\
    =&\sum_{\{s_i\}}\frac{d_1\cdots d_n}{\mathcal{D}^{2n}}
    \vcenter{\hbox{
    \begin{tikzpicture}[scale=.7]
        \foreach \x in {0,...,3}{
        \begin{scope}[shift={({\x*sqrt(3)},0)}]
            \draw[thick] (330:1) foreach \t in {30,90,...,330} { -- (\t:1) };
        \end{scope}
        }
        \filldraw[white] ({sqrt(3)*2-.5},-1.1) rectangle ++ (1,2.2);
        \node at ({sqrt(3)*2},0) {$\cdots$};
        \draw[very thick,violet,dashed] (0,0) circle (.4);
        \def\r{.65}
        \draw[very thick,violet,dashed,rounded corners=.2cm] (30:\r) foreach \t in {90,150,...,330} {-- (\t:\r)} foreach \t in {210,270,...,510} {--({cos(\t)*\r+sqrt(3)},{sin(\t)*\r})} -- cycle;
        \def\r{.85}
        \draw[very thick,violet,dashed,rounded corners=.17cm] (30:\r) foreach \t in {90,150,...,330} {-- (\t:\r)} foreach \t in {210,270,330} {--({cos(\t)*\r+sqrt(3)},{sin(\t)*\r})} foreach \t in {210,270,330} {--({cos(\t)*\r+2*sqrt(3)},{sin(\t)*\r})} foreach \t in {210,270,...,510} {--({cos(\t)*\r+3*sqrt(3)},{sin(\t)*\r})} foreach \t in {390,450,510} {--({cos(\t)*\r+2*sqrt(3)},{sin(\t)*\r})} foreach \t in {390,450,510} {--({cos(\t)*\r+1*sqrt(3)},{sin(\t)*\r})} -- cycle;

        \node[violet,font={\scriptsize}] at (.65,0) {$s_1$};
        \node[violet,font={\scriptsize}] at ({sqrt(3)+.3},0) {$s_2$};
        \node[violet,font={\scriptsize}] at ({3*sqrt(3)+.4},0) {$s_n$};
    \end{tikzpicture}}}
    \end{aligned}
\end{equation}

As a final note on the string net model, we mention that, as defined above, the model does not support excitations that violate only vertex operators. This is because the definition of the string operators \eqref{eq: string-op} assumes that the string around each vertex satisfies the fusion rules, yielding $W_s=0$ otherwise. 

This issue can be remedied by adding additional ``tails'' on the edges \cite{hu_full_2018,PhysRevX.12.021012}, with the ground state satisfying the additional requirement that the string on the tail is trivial
\begin{equation}
    S_q \ket{\vcenter{\hbox{
    \begin{tikzpicture}[scale=.5]
        \newdimen\r
    \r=1cm
    \draw[thick] (330:\r) foreach \x in {30,90,...,330} { -- (\x:\r) };
    \foreach \x in {30,90,...,330}{
        \draw[thick] (\x:\r) -- (\x:1.5*\r);
    }
    \draw[->,thick] ({sqrt(3)/2},0) -- ++ (.5,0) node[right] {$s$};
    \end{tikzpicture}}}}=
    \delta_{s,0} \ket{\vcenter{\hbox{
    \begin{tikzpicture}[scale=.5]
        \newdimen\r
    \r=1cm
    \draw[thick] (330:\r) foreach \x in {30,90,...,330} { -- (\x:\r) };
    \foreach \x in {30,90,...,330}{
        \draw[thick] (\x:\r) -- (\x:1.5*\r);
    }
    \draw[->,thick] ({sqrt(3)/2},0) -- ++ (.5,0) node[right] {$s$};
    \end{tikzpicture}}}},
\end{equation}
so that the Hamiltonian is
\begin{equation}
    H=-\sum_v A_v -\sum_p B_p -\sum_q S_q.
\end{equation}
This representation will be useful for us for creating the ground state of the string-net model. Notice that the addition of the tail adds two qudits to the edge, one additional qudit on the edge and another on the tail.

\subsection{A protocol for general string-nets}
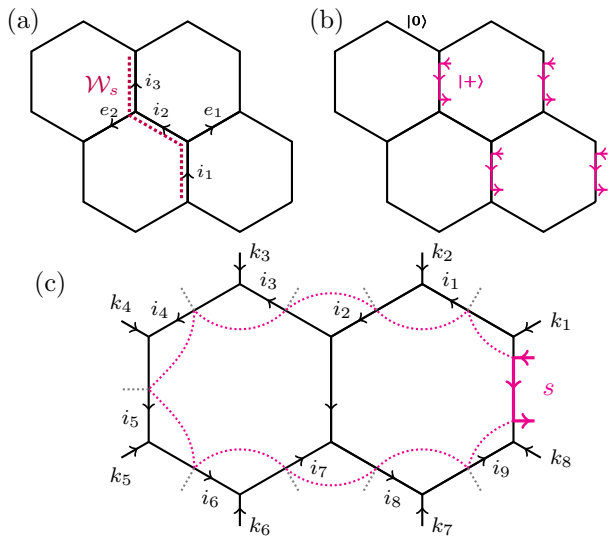
\begin{figure}
    \centering
    \begin{tikzpicture}[every node/.append style={font={\scriptsize}},scale=.8]
        \def\r{.3}
        \tikzset{string/.style={thick,midarr}}
        \def\ss{{sqrt(3)/2}}
        \foreach \x in {1,0}{
        \foreach \y in {1,0}{
        \begin{scope}[shift={({sqrt(3)*\x+sqrt(3)/2*\y},{-1.5*\y})}]
            \draw[thick] (330:1) foreach \t in {30,90,...,330} {-- (\t:1)};
        \end{scope}
        }}
        
        \draw[string] (\ss,-1/2) -- node[right] {$i_3$} (\ss,1/2);
        \draw[string] ({2*sqrt(3)/2},-2*1/2) -- node[above] {$i_2$} ({sqrt(3)/2},-1*1/2);
        \draw[string] ({2*sqrt(3)/2},-2*1/2) -- node[above] {$e_1$} ({3*sqrt(3)/2},-1*1/2);
        \draw[string] ({2*sqrt(3)/2},-4*1/2) -- node[right] {$i_1$} ({2*sqrt(3)/2},-2*1/2);
        \draw[string] ({1*sqrt(3)/2},-1*1/2) -- node[above] {$e_2$} ({0*sqrt(3)/2},-2*1/2);
        
        \def\sd{.12}
        \draw[purple,very thick,densely dotted] ({sqrt(3)/2*(2-\sd)},{1/2*(-4+\sd)}) -- ({sqrt(3)/2*(2-\sd)},{1/2*(-2-\sd)}) -- ({sqrt(3)/2*(1-1*\sd)},{1/2*(-1-\sd)})-- node[left] {\small $\mathcal{W}_s$} ({sqrt(3)/2*(1-1*\sd)},{1/2*(+1-\sd)});
        \node at (-1,1) {\normalsize (a)};
    \end{tikzpicture}
    \begin{tikzpicture}[scale=.8]
        \tikzset{string/.style={thick,magenta,arr={#1}}}
        \foreach \x in {1,0}{
        \foreach \y in {1,0}{
        \begin{scope}[shift={({sqrt(3)*\x+sqrt(3)/2*\y},{-1.5*\y})}]
            \draw[thick] (330:1) foreach \t in {30,90,...,330} {-- (\t:1)};
            \draw[string={.7}] ({sqrt(3)/2+.2},.3) -- ++ (-.2,0);
            \draw[string={.5}] ({sqrt(3)/2},.3) -- ++ (0,-.6);
            \draw[string={.8}] ({sqrt(3)/2},-.3) -- ++ (.2,0); 
        \end{scope}
        \node[magenta,font={\tiny}] at (1.4,0) {$\ket{+}$};
        \node[font={\tiny}] at (.5,1) {$\ket{0}$};
        }
        }
    \node at (-1,1) {\normalsize (b)};
    \end{tikzpicture}\\
    
    \begin{tikzpicture}[every node/.append style={font={\footnotesize}},scale=1.4]
        \tikzset{string/.style={very thick,magenta,arr={#1}}}
        \tikzset{bstring/.style={thick,arr={#1}}}
        \tikzset{ans/.style={thick,gray,densely dotted}}
        \def\ln{1.3};
        \def\s{1.73205081};
        \begin{scope}[shift={({sqrt(3)},0)}]
            \draw[thick] (330:1) foreach \t in {30,90,...,330} {-- (\t:1)};
            \draw[string={.7}] ({sqrt(3)/2+.2},.3) -- ++ (-.2,0);
            \draw[string={.5}] ({sqrt(3)/2},.3) -- ++ (0,-.6);
            \draw[string={.8}] ({sqrt(3)/2},-.3) -- ++ (.2,0); 
            \draw[bstring={.6}] (30:\ln) node[right] {$k_1$} -- (30:1);
            \draw[bstring={.6}] (90:\ln) node[right] {$k_2$} -- (90:1);
            \draw[bstring={.6}] (-90:\ln) node[right] {$k_7$} -- (-90:1);
            \draw[bstring={.6}] (-30:\ln) node[right] {$k_8$} -- (-30:1);
            \node[magenta] at (1.2,0) {\normalsize $s$};
            \foreach \t in {30,90,...,270}
                \draw[bstring={.7}] (\t:1) -- (\t+60:1);
            \foreach \t in {60,120,240,300}
                \draw[ans] (\t:{sqrt(3)/2}) -- (\t:{\ln*sqrt(3)/2});
            \node at (15+1*60:1.07) {$i_1$};
            \node at (15+2*60:1.07) {$i_2$};
            \node at (15+4*60:1.07) {$i_8$};
            \node at (15+5*60:1.07) {$i_9$};
        \end{scope}
        \foreach \t in {30,90,...,270}
            \draw[bstring={.7}] (\t:1) -- (\t+60:1);
        \foreach \t in {60,120,...,300}
            \draw[ans] (\t:{sqrt(3)/2}) -- (\t:{\ln*sqrt(3)/2});
        \draw[bstring={.6}] (90:\ln) node[right] {$k_3$} -- (90:1);
        \draw[bstring={.6}] (-90:\ln) node[right] {$k_6$} -- (-90:1);
        \draw[bstring={.6}] (150:\ln) node[above] {$k_4$} -- (150:1);
        \draw[bstring={.6}] (-150:\ln) node[below] {$k_5$} -- (-150:1);
        \node at (75:1.07) {$i_3$};
        \node at (15+60*2:1.07) {$i_4$};
        \node at (15+60*3:1.07) {$i_5$};
        \node at (15+60*4:1.07) {$i_6$};
        \node at (15+60*5:1.07) {$i_7$};
        \draw[magenta,thick,densely dotted] ({1.5*sqrt(3)},.3) to[out=160,in=-75] ($ (\s,0) + (60:\s/2) $) to[out=-135,in=-45] ($ (\s,0) + (120:\s/2) $) to[out=140,in=40] (60:\s/2) to[out=-135,in=-45] (120:\s/2) to[out=-90,in=45] (180:\s/2);
        \begin{scope}[yscale=-1]
        \draw[magenta,thick,densely dotted] ({1.5*sqrt(3)},.3) to[out=160,in=-75] ($ (\s,0) + (60:\s/2) $) to[out=-135,in=-45] ($ (\s,0) + (120:\s/2) $) to[out=140,in=40] (60:\s/2) to[out=-135,in=-45] (120:\s/2) to[out=-90,in=45] (180:\s/2);
        \end{scope}
    \node at (-1.8,1) {\normalsize (c)};
    \end{tikzpicture}
    \caption{Construction of a string-net model via loop closures. (a) the form of the string-net string operator. (b) The initial product state for the protocol generating a string net model. The black edges are initialized in the state $\ket{0}$ (the trivial string), while the purple edges are initialized in the state $\ket{+}_\C =\sum_s\frac{d_s}{\D}\ket{s}$. (c) A loop is closed by moving a string end around it, using the operator $\R$, see \eqref{eq:string-controlled-loop}.}
    \label{fig: string-net-creation}
\end{figure}
Let us describe a protocol similar to the one used in the $D(G)$ case above for creating the ground state of a general string-net model. This protocol is not FDSC but gives a basis for the protocols we describe in further sections that are FDSC. A similar method was described in \cite{liu2022methods}, with the difference that our description makes use of ``long loops'', which makes it useful as a basis for FDSC protocols.

As before, we break the hexagonal lattice into a spanning tree and its complement. Each edge on the spanning tree is initialized in the state $\ket{0}$ (with the trivial string). For edges at the complement of the spanning tree, we assign two additional qudits as tails on the edge. The initial product state on the edge is given by a superposition of all possible strings with appropriate weights: $\ket{+}_\C =\sum_s\frac{d_s}{\D}\ket{s}$, such that the strings start and end on the tails, and the three corresponding qudits are entangled (see Fig. \ref{fig: string-net-creation}b). The strings around the tails are initialized at $\ket{0}$. We now want to define operators that convert the short string on the edge of the complement to a string around some long loop closing the spanning tree. We note that the string operator \eqref{eq: string-op} creates two excitations on the edges of the string, but it is not guaranteed that the additional excitation fuses with the existing one to obtain the trivial string. Therefore, it cannot be used for the purpose of moving the string. However, we do have a unitary operator obtained from braiding, which is able to move the string end. Consider the operator acting on a small segment of the string net
\begin{equation}
\begin{aligned}
    \R&\ket{\vcenter{\hbox{
    \begin{tikzpicture}[scale=.7,every path/.append style={thick},every node/.append style={font={\scriptsize}}]
        \draw[midarr] (0,0) --  (-.5,0) node[below] {$i_3$};
        \draw[midarr] (1,0) -- node[below] {$i_2$} (0,0);
        \draw[midarr] (1.5,0) node[below] {$i_1$} --  (1,0);
        \draw[midarr] (0,1) -- node[right] {$k$} (0,0);
        \draw[midarr] (1,1) -- node[right] {$s$} (1,0);
    \end{tikzpicture}}}
    }\\
    =&\ket{\vcenter{\hbox{
    \begin{tikzpicture}[scale=.7,every path/.append style={thick},every node/.append style={font={\scriptsize}}]
        \draw[midarr] (0,0) --  (-.5,0) node[below] {$i_3$};
        \draw[midarr] (1,0) -- node[below] {$i_2$} (0,0);
        \draw[midarr] (1.5,0) node[below] {$i_1$} --  (1,0);
        \draw[arr={.8}] (1,1)  node[right] {$k$} -- (0,0);
        \filldraw[white] (.5,.5) circle (0.1);
        \draw[arr={.8}] (0,1) node[left] {$s$} -- (1,0);
    \end{tikzpicture}}}}\\
    =&\sum_{i_2'}\R_{i_1 s i_2}^{k i_3 i_2'}\ket{\vcenter{\hbox{
    \begin{tikzpicture}[scale=.7,every path/.append style={thick},every node/.append style={font={\scriptsize}}]
        \draw[midarr] (0,0) --  (-.5,0) node[below] {$i_3$};
        \draw[midarr] (1,0) -- node[below] {$i_2'$} (0,0);
        \draw[midarr] (1.5,0) node[below] {$i_1$} --  (1,0);
        \draw[midarr] (0,1) -- node[right] {$s$} (0,0);
        \draw[midarr] (1,1) -- node[right] {$k$} (1,0);
    \end{tikzpicture}}}} 
\end{aligned}
\end{equation}
with
\begin{equation}
    \R_{i_1 s i_2}^{k i_3 i_2'}=\sum_n F_{i_1 s \bar{n}}^{k\bar{i_3}i_2}R_{sk}^n F_{ksi_2'}^{\bar{i}_3i_1n},
\end{equation}
and $R$ obtained by a certain solution of the hexagon equations (the resulting unitary is independent of which solution is chosen). The unitary operator moving a string $s$ around a long loop of $l$ plaquettes on the spanning-tree edges can then be written as
\begin{equation}
    (\mathcal{L}^s_l)_{i_1...i_{4l+2}}^{i_1'...i_{4l+3}'}(k_0...k_{4l+1})=\prod_{t=0}^{4l+2}\R_{i_t' s i_t}^{k_t i_{t+1}i_{t+1}'},
\end{equation}
where we set $i_0=i_{4l+3}=1,i_0'=i_{4l+3}'=\bar{s}$ (see Fig. \ref{fig: string-net-creation}c). The ``controlled loop'' operation is then
\begin{equation}
    \label{eq:string-controlled-loop}
    CL_\ell = \sum_s \mathcal{L}^s_\ell P_s,
\end{equation}
where $P_s$ project on the rightmost edge of the loop having string $s$. 

The loop operators $CL_\ell$ can be interpreted as moving an anyon inside a loop, and as a result commute with each other. On the other hand, they are not, in general, FDSC operations, as the $\mathcal{R}$ operations have to be applied sequentially. It is therefore desirable to find cases in which the operators can be realized in FDSC operations.
\section{Abelian string nets}
\label{abelian-string-nets}
Abelian string-net models are the simplest string-net models that can be created by FDSC circuits. In this case, the fusion of two anyons only yields a single anyon, and as a result, the operators $W_s$ are unitary and, in fact
\begin{equation}
    W^s=\mathcal{L}_l^s
\end{equation}
on the relevant loops. Since the $F$-matrices are phases in the abelian case, from \eqref{eq: string-op} we see that the operators $W_s$ can be implemented as a multiplication of a set of commuting phase gates acting on the vertices, giving the $F$ phase factors, and a set of commuting gates that add a string $s$ on each gate.

\section{Non-abelian string nets: Gauging}
\label{string-nets-gauging}
We now show how non-abelian theories can be obtained by FDSC operations using the process of gauging an anyonic symmetry \cite{teo_theory_2015,barkeshli_symmetry_2019}. We start by defining the gauging procedure; we then construct Hamiltonians of non-abelian theories using this procedure; and then explain the FDSC process that creates the ground states of these Hamiltonians, in the cases where this is possible (similar to Sec. \eqref{quantum double}, we will find there the same condition of the group $G$ being solvable). 

The gauging procedure can be explained as follows: we assume that the theory $\C$ is symmetric under the action of a symmetry group $G$ which permutes the anyons via $a\mapsto g(a)$ for $g\in G$. By symmetry, we mean that the data describing the theory ($N$ and $F$ matrices) is invariant under the transformation. In general, this symmetry can be a true global symmetry of the model (as will be in the following discussion) or a symmetry of the low energy theory describing the ground state and anyonic excitations. We want to gauge the symmetry, that is, make the global symmetry a local symmetry of the ground state. This is done by adding a gauge field between the degrees of freedom in the model, such that the resulting local symmetry action is a combination of a change in the anyon label and an appropriate change in the gauge field. Gauge invariance in this case will be enforced as an energetic term on the ground state.

\subsection{The gauged string-net model}
Let us describe the gauging procedure, with the goal of defining the ground state that will later be created using an FDSC circuit. For a given string-net model built from a theory $\C$, we define a new lattice model describing the gauged theory. Similarly to the string-net model, the gauged string-net is defined on the honeycomb lattice, with edge qudits obtaining values in the string labels. In addition, we put around each vertex three qudits connecting the three edges in the vertex (see Fig. \ref{fig:gauged-model}). We want the ground state to be invariant under gauge transformations of the form
\begin{equation}
    \L^g_e\ket{\vcenter{\hbox{
    \begin{tikzpicture}[every node/.append style={font={\scriptsize}},scale=1.2]
        \draw[thick,midarr,orange] (1,0) ++(60:.3) arc (60:180:.3) node[midway,above] {$g_1$};
        \draw[thick,midarr,orange] (0,0) ++(0:.3) arc (0:120:.3) node[midway,above] {$g_2$};
        \draw[thick,midarr,orange] (0,0) ++(-120:.3) arc (-120:0:.3) node[midway,below] {$g_3$};
        \draw[thick,midarr,orange] (1,0) ++(-180:.3) arc (-180:-60:.3) node[midway,below] {$g_4$};
        \draw[very thick,midarr] (0,0) -- node[above] {$s$} (1,0);
    \end{tikzpicture}
    }}}=
    \ket{\vcenter{\hbox{
    \begin{tikzpicture}[every node/.append style={font={\scriptsize}},scale=1.2]
        \draw[thick,midarr,orange] (1,0) ++(60:.3) arc (60:180:.3) node[midway,above] {$g_1g^{-1}$};
        \draw[thick,midarr,orange] (0,0) ++(0:.3) arc (0:120:.3) node[midway,above] {$gg_2$};
        \draw[thick,midarr,orange] (0,0) ++(-120:.3) arc (-120:0:.3) node[midway,below] {$g_3g^{-1}$};
        \draw[thick,midarr,orange] (1,0) ++(-180:.3) arc (-180:-60:.3) node[midway,below] {$gg_4$};
        \draw[very thick,midarr] (0,0) -- node[above] {$g(s)$} (1,0);
    \end{tikzpicture}
    }}}.
\end{equation}
This is ensured when the ground state is an eigenstate of the operators acting on the string-net edges
\begin{equation}
    \L_e=\frac{1}{|G|}\sum_{g\in G}\L^g_e.
\end{equation}
with eigenvalue 1. The ground state should also have zero flux at any gauge field loop, enforced by the projections
\begin{align}
    \Pi_v\ket{\vcenter{\hbox{
    \begin{tikzpicture}[scale=.8,every node/.style={font=\tiny}]
        \foreach \q in {1,2,3}{
            \draw[thick,orange,midarr] (\q*120+210:.4) arc (\q*120+210:\q*120+330:.4);
            \draw[thick] (0,0) -- (\q*120+90:1);
            \node[orange] at (270+120*\q:.7) {$g_{\q}$};
        }
    \end{tikzpicture}
    }}}&=\delta_{g_1g_2g_3} \ket{\vcenter{\hbox{
    \begin{tikzpicture}[scale=.8,every node/.style={font=\tiny}]
        \foreach \q in {1,2,3}{
            \draw[thick,orange,midarr] (\q*120+210:.4) arc (\q*120+210:\q*120+330:.4);
            \draw[thick] (0,0) -- (\q*120+90:1);
            \node[orange] at (270+120*\q:.7) {$g_{\q}$};
        }
    \end{tikzpicture}
    }}},\\
    \Pi_p\ket{\vcenter{\hbox{
    \begin{tikzpicture}[scale=.8]
        \foreach \q in {1,...,6}{
            \def\r{.3}
            \draw[thick,orange,midarr] (-\q*60-30:1) ++ (-\q*60+90:\r) arc (-\q*60+90:-\q*60+210:\r);
            \node[orange] at (-\q*60-30:1-\r-.2) {\scriptsize $g_\q$};
            \draw[thick] (\q*60-30:1) -- (\q*60+30:1);
        }    
    \end{tikzpicture}
    }}}    
    &=\delta_{g_1g_2g_3g_4g_5g_6}\ket{\vcenter{\hbox{
    \begin{tikzpicture}[scale=.8]
        \foreach \q in {1,...,6}{
            \def\r{.3}
            \draw[thick,orange,midarr] (-\q*60-30:1) ++ (-\q*60+90:\r) arc (-\q*60+90:-\q*60+210:\r);
            \node[orange] at (-\q*60-30:1-\r-.2) {\scriptsize $g_\q$};
            \draw[thick] (\q*60-30:1) -- (\q*60+30:1);
        }    
    \end{tikzpicture}
    }}}.
\end{align}

This is sufficient for defining the state of the gauge field. We notice that in the case where the string model is trivial (only the trivial string, and a one-dimensional Hilbert space on the honeycomb lattice) the resulting model is exactly the Kitaev quantum double model discussed in Sec. \ref{quantum double}. This represents the fact that the quantum double model is the gauging of the trivial $G$ symmetry of a trivial insulator.

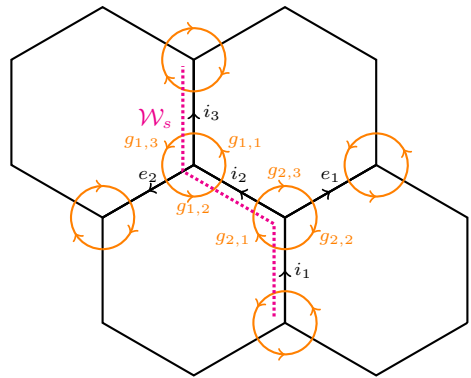
\begin{figure}
    \centering
    \begin{tikzpicture}[every node/.append style={font={\scriptsize}},scale=1.4]
        \def\r{.3}
        \tikzset{string/.style={thick,midarr}}
        \def\ss{{sqrt(3)/2}}
        \foreach \x in {1,0}{
        \foreach \y in {1,0}{
        \begin{scope}[shift={({sqrt(3)*\x+sqrt(3)/2*\y},{-1.5*\y})}]
            \draw[thick] (330:1) foreach \t in {30,90,...,330} {-- (\t:1)};
        \end{scope}
        }}
        
        \draw[string] (\ss,-1/2) -- node[right] {$i_3$} (\ss,1/2);
        \draw[string] ({2*sqrt(3)/2},-2*1/2) -- node[above] {$i_2$} ({sqrt(3)/2},-1*1/2);
        \draw[string] ({2*sqrt(3)/2},-2*1/2) -- node[above] {$e_1$} ({3*sqrt(3)/2},-1*1/2);
        \draw[string] ({2*sqrt(3)/2},-4*1/2) -- node[right] {$i_1$} ({2*sqrt(3)/2},-2*1/2);
        \draw[string] ({1*sqrt(3)/2},-1*1/2) -- node[above] {$e_2$} ({0*sqrt(3)/2},-2*1/2);

        \foreach \vertex in {({sqrt(3)/2*1},{1/2*1}),({sqrt(3)/2*2},{-1/2*2}),({sqrt(3)/2*0},{-1/2*2})}{
        \foreach \q in {30,150,270}
            \draw[thick, orange,midarr] \vertex ++ (\q:\r) arc (\q:\q-120:\r);
        }
        
        \foreach \vertex in {({sqrt(3)/2*1},{-1/2*1}),({sqrt(3)/2*3},{-1/2*1}),({sqrt(3)/2*2},{-1/2*4})}{
        \foreach \q in {-30,90,210}
            \draw[thick, orange,midarr] \vertex ++ (\q:\r) arc (\q:\q+120:\r);
        }

        \foreach \q in {1,2,3}{
            \node[orange] at ({sqrt(3)+1.9*\r*cos(120*\q+90)},{-1+1.4*\r*sin(120*\q+90)}) {$g_{2,\q}$};
            \node[orange] at ({sqrt(3)/2+1.9*\r*cos(-120*\q-210)},{-1/2+1.4*\r*sin(-120*\q-210)}) {$g_{1,\q}$};
            }

        \def\sd{.12}
        \draw[magenta,very thick,densely dotted] ({sqrt(3)/2*(2-\sd)},{1/2*(-4+\sd)}) -- ({sqrt(3)/2*(2-\sd)},{1/2*(-2-\sd)}) -- ({sqrt(3)/2*(1-1*\sd)},{1/2*(-1-\sd)})-- node[left] {\small $\mathcal{W}_s$} ({sqrt(3)/2*(1-1*\sd)},{1/2*(+1-\sd)});
    \end{tikzpicture}

    \caption{The gauged string-net model. The model is defined with edges of the hexagonal lattice labels by a string $s$, and "gauge field" edges connecting any two edges along a vertex, and obtaining values in the gauge group $G$. Also presented is the definition of a string operator $\mathcal{W}^s$, which adds a string $s$ modified by the appropriate action of the gauge group.}
    \label{fig:gauged-model}
\end{figure}

Next, we want to define the interaction of the strings with the gauge field such that the string projectors are compatible with the $\L^g_e$ actions. The vertex operators are defined as
\begin{equation}
    \mathcal{A}_v\ket{\vcenter{\hbox{
    \begin{tikzpicture}[scale=.8,every node/.style={font=\tiny}]
        \foreach \q in {1,2,3}{
            \draw[thick,orange,midarr] (\q*120+210:.4) arc (\q*120+210:\q*120+330:.4);
            \node[orange] at (270+120*\q:.7) {$g_{\q}$};
        }
    \draw[thick,midarr] (210:1) node[above] {\footnotesize$a$} -- (0,0);
    \draw[thick,midarr] (330:1) node[above] {\footnotesize$b$} -- (0,0);
    \draw[thick,arr={.7}] (0,0)  -- (0,1) node[right] {\footnotesize$c$};
    \end{tikzpicture}
    }}} = N_{g_2(a),g_2g_3(b)}^{c}\ket{\vcenter{\hbox{
    \begin{tikzpicture}[scale=.8,every node/.style={font=\tiny}]
        \foreach \q in {1,2,3}{
            \draw[thick,orange,midarr] (\q*120+210:.4) arc (\q*120+210:\q*120+330:.4);
            \node[orange] at (270+120*\q:.7) {$g_{\q}$};
        }
    \draw[thick,midarr] (210:1) node[above] {\footnotesize$a$} -- (0,0);
    \draw[thick,midarr] (330:1) node[above] {\footnotesize$b$} -- (0,0);
    \draw[thick,arr={.7}] (0,0)  -- (0,1) node[right] {\footnotesize$c$};
    \end{tikzpicture}
    }}}
\end{equation},
and we see that indeed $[\mathcal{A}_v,\L_e]=0$ on the ground states of $\Pi_v$.

To define the string plaquette operators, we begin with the gauged version of the operators $W_s$ that add a string $s$ to a path of edges. Since the gauge field keeps track of the $g$ action over a path of edges, the string operator should be compatible with it. This means that if we add a string $s$ on an edge, we should add a string $g(s)$ on the next edge, with $g$ being the value of the gauge field between the edges. Explicitly, this gives
\begin{align}
    \mathcal{W}_{i_1,i_2,...,i_N}^{i_1',i_2',...,i_N'}&=\prod_{k=1}^N F_{\bar{s},g^{(k)}(i_{k}'),g^{(k+1)}(\bar{i}_{k+1}')}^{g_{k,3}(e_k),g^{(k+1)}(\bar{i}_{k+1}),g^{(k)}(i_{k})}, \\
    g^{(k)}&=g_{k-1,1}\cdots g_{1,1}
\end{align}
(see Fig \ref{fig:gauged-model}). While this expression becomes quite intimidating, it simply modifies \eqref{eq: string-op} so that we add to each edge the appropriate string, gauged by the appropriate group action. It can be seen as the operator $W_s$ conjugated by gauge transformations $\L_v$ that trivialize the gauge field around the vertices where the string passes. The string plaquette operators are then obtained by summing over all closed string operators on that plaquette
\begin{equation}
    \mathcal{B}_p = \sum_{s} \frac{d_s}{\DD^2} \mathcal{W}_p^s,
\end{equation}
with $d_s,\DD$ as defined for $\C$. Defining all projection operators of the model, we can write the Hamiltonian as 
\begin{equation}
\label{eq:gauged-H}
    H_{\rm gauged} = -\sum_e \L_e-\sum_v\qty(\Pi_v+\mathcal{A}_v)-\sum_p \qty(\mathcal{B}_p+\Pi_p).
\end{equation}
The model as defined here yields the theory obtained from gauging the symmetry $G$ in its simplest form. We give a brief discussion of the resulting gauged theory in Appendix \ref{app:gauged theory}. In general, we can allow for the symmetry action on $\C$ to be fractionalized (that is, each anyon carries a fractional ``$G$ charge'') and the symmetry action can be twisted by a non-trivial element of the cohomology $H^3(G, U(1))$ \cite{barkeshli_symmetry_2019,levin2012braiding}. We will not describe these more complicated constructions here.

\subsection{FDSC circuit for creating the gauged string-net state}
\begin{figure}[b]
    \centering
    \begin{tikzpicture}[scale=1]
        \newdimen\r
    \draw[thick] (330:1) foreach \x in {30,90,...,330} { -- (\x:1) };
    \foreach \x in {90,150,...,270}{
        \draw[thick] (\x:1) -- (\x:1.5);
    }
    \draw[thick] (30:1) -- (30:2);
    \draw[thick,arr=.7] (330:1) -- (330:2);
    \node at (2.5,-.5) {$s\mapsto g_3 g_2 g_1 s$};
    \draw[thick,orange,midarr] (210:1) ++ (-30:.4) arc (-30:90:.4);
    \node[orange,font={\tiny}] at (210:.4) {$g_1$};
    \draw[thick,orange,midarr] (270:1) ++ (30:.4) arc (30:150:.4);
    \node[orange,font={\tiny}] at (270:.4) {$g_2$};
    \draw[thick,orange,midarr] (330:1) ++ (210:.4) arc (210:330:.4);
    \node[orange,font={\tiny}] at ({sqrt(3)/2},-1.2) {$g_3$};
    \draw[dashed,internationalkleinblue,thick] plot [smooth] coordinates {(180:{sqrt(3)/2})  (240:{sqrt(3)/2}) (300:{sqrt(3)/2}) (330:1.5)};
    \filldraw[internationalkleinblue] (180:{sqrt(3)/2}) circle (.07);
    \filldraw[internationalkleinblue] (330:1.5) circle (.07);
    \end{tikzpicture}\\
    \caption{Obtaining the gauged model by acting on each string by $g_1,..,g_n$ where the $g_i$'s are obtained from the path leading to it. }
    \label{fig:gauging}
\end{figure}
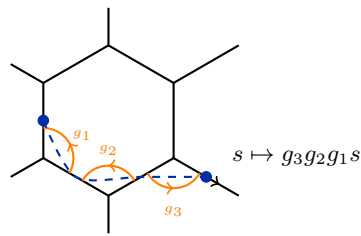

Once the model is defined, the methods introduced above can be used to create the ground state of that Hamiltonian \eqref{eq:gauged-H}. To do so, we first create the ground state of the string net model $\C$ on the string qubits, and, independently, the ground state of the quantum double $D(G)$ on the gauge field qubits. Next, we choose an edge $e_0$ on the string lattice and for each other edge of the graph we pick a path going to it from $e_0$. Finally, we act on the edge with an operator that sends $s\mapsto g_0\cdots g_n (s)$ where $g_0,...,g_n$ are the values of the gauge fields connecting the two edges on the chosen path (see Fig. \ref{fig:gauging}). From the zero flux condition, the product $g_0\cdots g_n$ is independent of the choice of the path. Furthermore, if $G$ is solvable the operator can be applied by simultaneous commuting gates of finite depth using the results of Sec. \ref{quantum double}. Together, this shows that if $\C$ can be created in finite depth and $G$ is solvable, the gauged theory $\C^G$ can also be created in finite depth. Note that this procedure requires additional ancillas only for the calculation of the products $g_0\cdots g_n$. In that case, similar to the quantum double case, the required number of ancillas is proportional to the system size. This is because one can put ancillas on the edges storing the values of products of elements leading to it (see Eq. \eqref{eq:solvable-product}). In the case where $G$ is abelian, no additional ancillas are required.

\section{General CSS codes}
\label{css-codes}
Our previous sections discussed the creation of 2+1d TO states. Since the computational model we consider is free of geometric constraints on the circuits, more general quantum codes, not necessarily defined on a planar graph, can similarly be realized. We now generalize the discussion of the toric code in Sec. \ref{toric code} to general CSS codes (see below for a definition of such codes). Our discussion will be rather abstract, presenting a more algebraic description of our prescription. Our goal is to show how for a general CSS code one can define an FDSC circuit which gives a ground state of the code, and describe how, in general, such a circuit can be optimized in terms of the number of two-qubit gates. We present a polynomial-time algorithm for finding an FDSC circuit given a CSS code, but we note that finding an optimal circuit will in general require solving an optimization problem over the integers, and thus is expected to require exponential time in the number of qubits.

A general CSS code is defined on a set of $N$ qubits, with the Hamiltonian
\begin{equation}
    H=-\sum_{i\in J_A} A_i -\sum_{j\in J_B} B_j,
\end{equation}
such that $A_i$ acts as a product of $X$ operators on different qubits, $B_j$ acts as a product of $Z$ operators, and $[A_i,B_j]=0$. The sets $J_A,J_B$ are lists of indices for $A_i,B_j$. A ground state of the model can be obtained via
\begin{equation}
\label{eq:psi-css}
\begin{aligned}
    \ket{\y}&\propto \prod \frac{1+A_i}{2}\ket{0^N} \\
    &\propto\sum_{\{\Phi_i\}}\prod_{i\in I}\qty[(1-\Phi_i)I+\Phi_iA_i ]\ket{0^N},
\end{aligned}
\end{equation}
where $\Phi_i$ sums over values in 0,1 (generally, this ground state is not unique). The ground state is then obtained as a superposition indexed by values of $\Phi$. Note that it is independent of the form of the $B_j$ operators (as long as they commute with $A_i$), but the exact form of $B_j$ does determine the ground state degeneracy and the form of the other ground states.

Our strategy will be as follows:
\begin{enumerate}
    \item Find a subset $S$ of qubits which encode the entire state of the code qubits. That is, for a given code ground state, if the $S$ qubits are measured in the $Z$ basis, the $Z$ values of all qubits can be inferred (in the toric code example described above, this is given by the spanning tree).
    \item Obtain a linear (mod 2) function $\MM_S$ relating the values of the $S$ qubits (in the $Z$ basis) to the values of all qubits in the code.
    \item The protocol is then given by starting with the qubits in $S$ in the $\ket{+}$ state, all other qubits in the $\ket{0}$ state, then applying a set of $CX_{i,j}$ operations, with $i,j$ obtained from the nonzero entries of $\MM_S$. The total number of $CX$ operations is then given by the number of non-zero entries in $\MM_S$.
\end{enumerate}

To describe the elements of the superposition \eqref{eq:psi-css} it is useful to work with linear algebra in $\ft$ (the field of two elements). Each element in the superposition is given by a set of values $z_i$ of the qubits, with $z_i\in\{0,1\}$. The qubit configuration obtained from given values of ${\Phi_i}$ is
\begin{equation}
    \vec{z} = \AA\vec{\Phi},
\end{equation}
where $\AA\in \ft^{N\times\abs{J_A}}$ has each column obtained from $A_i$. The number of elements in the superposition is given by $2^{\abs{\im{\AA}}}$ (the dimension of the image of $\AA$). Notice that this means that the description in terms of $\AA$ is redundant: $\vec{\Phi}$ is an $\abs{J_A}=\abs{\ker \AA}+\abs{\im A}$ dimensional vector, so multiple values of $\vec{\Phi}$ give the same $\vec{z}$.

For each subset $S$ of qubits, consider the operator $\pi_S\in \ft^{\abs{S}\times N}$ which projects to the set $S$. To carry out our protocol, we want to find a subset $S$ for which the $z$ values can be arbitrary in the superposition comprising $\ket{\psi}$, but whose values encode the values of all $z$ in the state. The first requirement translated in the toric code example to the requirement that the subset has no loops (i.e. it is a tree). Algebraically, it means that 
\begin{equation}
\label{eq:s-dim}
    \abs{S}=\abs{\im(\pi_S \AA)}
\end{equation}
(since any value of $S$ can be obtained by projecting an output of $\AA$). As for the second, we want to be able to reconstruct $\vec{z}$ from the projection $\pi_S \vec{z}$. A sufficient condition is that
\begin{equation}
\label{eq:s-dim-2}
    \abs{S} = \abs{\im \AA}
\end{equation}
(for the toric code, this is the requirement that the tree is \textit{spanning}). This is because we want the number of configurations in the superposition to be the number of possible configurations of $S$ qubits. 

When the two requirements above are satisfied, we can obtain the relation between the values of qubits in $S$ and the full set of qubits. The requirement \eqref{eq:s-dim} allows the definition of a right inverse (e.g. the Moore-Penrose pseudoinverse) \footnote{In general, the right inverse is not unique. However, any other choice will give the same operator $\MM_S$ in \eqref{eq:m-op-defn}.}
\begin{equation}
    (\pi_S\AA)^+=\AA^T\pi_S^T \qty(\pi_S\AA\AA^T\pi_S^T)^{-1}.
\end{equation}
We can then define an operator from $S$ to the full set of qubits given by 
\begin{equation}
    \label{eq:m-op-defn}
    \MM_S=\AA(\pi_S \AA)^+.
\end{equation}
The operator $\MM_S$ is an $N\times \abs{S}$ matrix which reconstructs the values $\vec{z}$. Notice that the rows of $\MM_S$ corresponding to qubits $i\in S$ will simply be 1 for the $i$'th entry and 0 otherwise. The condition \eqref{eq:s-dim-2} then ensures that $\abs{\im \MM_S}=\abs{\im \AA}$ so $\im{\MM_S}=\im{\AA}$. That is, the full $\vec{z}$ values of the code can be reconstructed from $\pi_S\vec{z}$. 

Given a subset $S$ satisfying \eqref{eq:s-dim} and \eqref{eq:s-dim-2}, we are now in a position to describe the FDSC circuit obtaining $\ket{\psi}$ as in \eqref{eq:psi-css}. We start with a product state of $\ket{+}$ on the $S$ qubits and $\ket{0}$ on all other qubits. We then apply the unitary
\begin{equation}
\label{eq: U_css}
    U_S=\prod_{i\in S,j=\{1,...,N\}} \begin{cases}
        CX_{i,j},& \qty(\MM_S)_{i,j}=1 \textrm{ and } i\neq j \\
        I,&{\rm otherwise}
    \end{cases}.
\end{equation}
This ensures that the resulting state is given by \eqref{eq:psi-css}. The total number of $CX$ operations is then given by $\norm{\MM_S}-\abs{S}$ where $\norm{\MM_S}$ is the number of nonzero elements in $\MM_S$. In the toric code case presented in \ref{toric code} where the set $S$ is given by the spanning tree, the operator $U_S$ is then exactly $\utc$ obtained in \eqref{eq:tc-unitary}.

The only remaining question is how $S$ can be found. This can be done by the following greedy algorithm: Start with the empty set $S_0$. In each step $t$ pick a random qubit $r_t$. If \eqref{eq:s-dim} is satisfied for the set $S_{t-1}\cup\qty{r_t}$ we let $S_t=S_{t-1}\cup\qty{r_t}$, otherwise we pick a different random $r_t$ and repeat until success. We stop when $t=\abs{\im \AA}$ and set $S=S_\abs{\im \AA}$. In Appendix \ref{app:algorithm} we show that for $t<\abs{\im \AA}$ an $r_t$ satisfying the required condition can always be found so that the algorithm is guaranteed to converge. The algorithm can be repeated multiple times to find $S$ with a small value of $\norm{\MM_S}$. By definition of \eqref{eq: U_css}, the algorithm is guaranteed to obtain a circuit using $O(N^2)$ gates (with $N$ being the number of qubits). This scaling is not necessarily optimal, however finding the optimal number of gates (even up to a constant) is generally expected to be a computationally hard problem.

\section{3D fracton models: the X-Cube model and Haah's Code}
\label{fractons}
Following the discussion in the previous section, we consider the more concrete example of 3D fracton models. In this section, we give two examples of the usage of the formalism developed above: the first is the X-Cube model \cite{vijay2016fracton}, for which we give a protocol requiring $O(L^4)$ two-qubit gates. The second example we give is Haah's code. Focusing on open boundary conditions, we present a protocol for obtaining a ground state of the model. We present numerical evidence showing that the protocol requires $L^5$ two-qubit gates (possibly up to logarithmic factors).
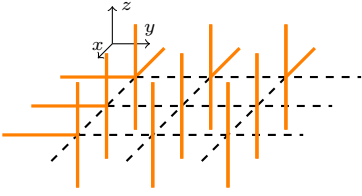
\begin{figure}
    \centering
        
    \begin{tikzpicture}[every node/.style={font=\scriptsize}]
        \def\s{3};
        \def\v{.7}
        \draw[->,thin] (.5,.25,.5) -- ++ (.5,0,0) node[above] {$y$};
        \draw[->,thin] (.5,.25,.5) -- ++ (0,.5,0) node[right] {$z$};
        \draw[->,thin] (.5,.25,.5) -- ++ (0,0,.5) node[above] {$x$};
        \foreach \x in {1,...,\s}{
        \draw[thick,dashed] (\x,0,1) -- (\x,0,\s+1);
        \draw[very thick,orange] (\x,0,0) -- (\x,0,1);
        \draw[thick,dashed] (1,0,\x) -- (\s+1,0,\x);
        \draw[very thick,orange] (0,0,\x) -- (1,0,\x);
        }
        \foreach \x in {1,...,\s}{
        \foreach \z in {1,...,\s}{
        \draw[very thick,orange] (\x,-\v,\z) -- (\x,\v,\z);
        \draw[very thick,orange] (\x,-\v,\z) -- (\x,\v,\z);
        }
        }
    \end{tikzpicture}
    \caption{The initial configuration for the circuit generating the X-Cube model. The vertical edges, as well as the horizontal edges on the farthest left and back (orange in the figure), are initialized in the $\ket{+}$ state. The other horizontal edges (dashed in the figure) are initialized in the $\ket{0}$ state.}
    \label{fig:x-cube-init}
\end{figure}
\subsection{X-Cube model}
The X-Cube model is a fracton model closely related to the 2+1d toric code. As such, our protocol above for obtaining the toric code works with little modification for obtaining the ground state of the X-Cube model. We begin by presenting the model and then show the FDSC circuit generating it.
The X-Cube model is defined on the 3D square lattice with a single qubit on each edge. The Hamiltonian has a cube operator and three types of vertex operators. They are given by
\begin{align}
    B_c &= \vcenter{\hbox{\begin{tikzpicture}[every path/.append style={thick},every node/.style={font={\tiny}}]
        \draw (0,0,0) -- node[below] {$X$} (1,0,0) -- node[right] {$X$} (1,1,0)-- node[above] {$X$} (0,1,0) -- node[left] {$X$} cycle;
        \draw (0,0,1) -- node[below] {$X$} (1,0,1) -- node[right] {$X$} (1,1,1)-- node[above] {$X$} (0,1,1) -- node[left] {$X$} cycle;
        \draw (0,0,0) -- node[right] {$X$} (0,0,1);
        \draw (1,0,0) -- node[right] {$X$} (1,0,1);
        \draw (0,1,0) -- node[above] {$X$} (0,1,1);
        \draw (1,1,0) -- node[below] {$X$} (1,1,1);
    \end{tikzpicture}}}, \\
    A_{v,x}&=\vcenter{\hbox{
    \begin{tikzpicture}[every path/.append style={thick},every node/.style={font={\tiny}},scale=.6]
        \draw(0,-1,0) -- node[right] {$Z$} (0,0,0) -- node[left] {$Z$} (0,1,0);
        \draw(0,0,-1) -- node[right] {$Z$} (0,0,0) -- node[left] {$Z$} (0,0,1);
        \draw[dotted] (-1,0,0) -- (1,0,0);
    \end{tikzpicture}}};\ 
    A_{v,y}=\vcenter{\hbox{
    \begin{tikzpicture}[every path/.append style={thick},every node/.style={font={\tiny}},scale=.6]
        \draw(-1,0,0) -- node[above] {$Z$} (0,0,0) -- node[below] {$Z$} (1,0,0);
        \draw(0,0,-1) -- node[above] {$Z$} (0,0,0) -- node[below] {$Z$} (0,0,1);
        \draw[dotted] (0,-1,0) -- (0,1,0);
    \end{tikzpicture}
    }};\ 
    A_{v,z}=\vcenter{\hbox{
    \begin{tikzpicture}[every path/.append style={thick},every node/.style={font={\tiny}},scale=.6]
        \draw(-1,0,0) -- node[fill=white] {$Z$} (0,0,0) -- node[fill=white] {$Z$} (1,0,0);
        \draw(0,-1,0) -- node[fill=white] {$Z$} (0,0,0) -- node[fill=white] {$Z$} (0,1,0);
        \draw[dotted] (0,0,-1) -- (0,0,1);
    \end{tikzpicture}
    }},
\end{align}
such that the Hamiltonian is
\begin{equation}
    H_{\rm X-cube} = -\sum_c B_c - \sum_{v,l} A_{v,l}.
\end{equation}
Our circuit for obtaining the ground state of the X-Cube model is a generalization of the protocol for obtaining the toric code. We start with the state depicted in Fig. \ref{fig:x-cube-init}, where all vertical qubits, as well as two planes of the horizontal qubits (those crossing the $xz$ and $yz$ planes), are initialized in the $\ket{+}$ state, and the other qubits are initialized in $\ket{0}$. The choice of the states initialized with $\ket{+}$ should be understood as follows: for each $xy,yz,xz$ plane the edges are a spanning tree of the \textit{dual} lattice on that plane.

The entangling circuit is composed of operators closing ``loops", but now the loops are in the dual lattice of each plane, that is
\begin{align}
    U_{\rm x-cube} &= \prod_\ell CL^{\rm x-cube}_\ell, \\
\label{eq:x-cube-cl}
CL_\ell&=\vcenter{\hbox{
    \begin{tikzpicture}[scale=.87]
    \draw[thick,orange] (0,0) -- (1,0);
    \foreach \x in {1,...,3}{
        \draw[very thick,orange] (\x,-1) --   (\x,1);
        \draw[thick,dashed] (\x,0) -- (\x+1,0);
        \filldraw (\x,.5) circle (.05);
        \filldraw (\x,-.5) circle (.05);
        \draw[thin] (\x,.5) -- (3.5,0);
        \draw[thin] (\x,-.5) -- (3.5,0);
        }
    \filldraw (.5,0) circle (.05);
    \draw (.5,0) to[in=170,out=10] (3.5,0);
    \begin{scope}[shift={(3.5,0)}]
    \filldraw[fill=white] (0,0) circle (.2);
    \draw (-.2,0) -- (.2,0);
    \draw (0,-.2) -- (0,.2);
    \end{scope}
\end{tikzpicture}}}.
\end{align}
The protocol discussed above requires $O(L^4)$ gates. We conjecture that, similar to the toric code, the required number of gates can be brought down to $O(L^3)$ by optimizing the choice of qubits initialized as $\ket{+}$. 

\subsection{Haah's code}
Haah's cubic code \cite{haah2011local} is another fracton model of interest, being the most prominent example of a code state without mobile excitations. The state is defined on a cubic lattice with two qubits placed at each vertex. The two types of stabilizers of the model are given by
\begin{align}
\label{eq:haah-stabilisers}
    A_{c,Z}&=\vcenter{\hbox{
    \begin{tikzpicture}[every node/.style={font={\tiny},fill=white},every path/.append style={thick}]
        \draw[dotted] (0,0,0) -- (1,0,0);
        \draw[dotted] (0,0,0) -- (0,1,0);
        \draw[dotted] (0,0,0) -- (0,0,1);
        \draw (1,0,0) -- (1,1,0) -- (1,1,1) -- (1,0,1) -- cycle;
        \draw (0,1,0) -- (1,1,0) -- (1,1,1) -- (0,1,1) --cycle;
        \draw (0,0,1) -- (1,0,1) -- (1,1,1) -- (0,1,1) -- cycle;
        \node at (0,0,0) {$II$};
        \node at (1,0,0) {$IZ$};
        \node at (0,1,0) {$IZ$};
        \node at (0,0,1) {$IZ$};
        \node at (1,1,0) {$ZI$};
        \node at (1,0,1) {$ZI$};
        \node at (0,1,1) {$ZI$};
        \node at (1,1,1) {$ZZ$};
    \end{tikzpicture}
    }},&
    A_{c,X}&=\vcenter{\hbox{
    \begin{tikzpicture}[every node/.style={font={\tiny},fill=white},every path/.append style={thick}]
        \draw[dotted] (0,0,0) -- (1,0,0);
        \draw[dotted] (0,0,0) -- (0,1,0);
        \draw[dotted] (0,0,0) -- (0,0,1);
        \draw (1,0,0) -- (1,1,0) -- (1,1,1) -- (1,0,1) -- cycle;
        \draw (0,1,0) -- (1,1,0) -- (1,1,1) -- (0,1,1) --cycle;
        \draw (0,0,1) -- (1,0,1) -- (1,1,1) -- (0,1,1) -- cycle;
        \node at (0,0,0) {$XX$};
        \node at (1,0,0) {$IX$};
        \node at (0,1,0) {$IX$};
        \node at (0,0,1) {$IX$};
        \node at (1,1,0) {$XI$};
        \node at (1,0,1) {$XI$};
        \node at (0,1,1) {$XI$};
        \node at (1,1,1) {$II$};
    \end{tikzpicture}
    }},
\end{align}
(one can check that $[A_{c,x},A_{c',z}]=0$). Here we consider the model with ``open" boundary conditions (BC), such that the ground state is obtained in the form of \eqref{eq:psi-css} where $A_i$ are $A_{c,X}$ as defined in \eqref{eq:haah-stabilisers} on each one of $L\times L\times L$ cubes (this defines the model on $2(L+1)^3$ qubits). The case of periodic BC is very interesting in Haah's code, as it leads to nontrivial dependence of the ground-state degeneracy on the system size \cite{haah2011local}. While the application of the methods of Sec. \ref{css-codes} is straightforward also in the periodic BC, it cannot be done analytically, and we focus on open BC here for a more conceptual presentation. 

Each element in the superposition \eqref{eq:psi-css} is defined by values of the $Z$ operators on each qubit, as $z_{(x,y,z;i)}$. They are related to the values of the ``potential" $\Phi$ by
\begin{align}
\label{eq:z1-by-phi}
    z_{xyz;1}&=\Phi_{x^+y^+z^+}\Phi_{x^+y^-z^-}\Phi_{x^-y^+z^-}\Phi_{x^-y^-z^+}, \\
    \label{eq:z2-by-phi}
    z_{xyz;2}&=\Phi_{x^+y^+z^+}\Phi_{x^-y^+z^+}\Phi_{x^+y^-z^+}\Phi_{x^+y^+z^-},
\end{align}
where we use e.g. $x^\pm\equiv x\pm\frac{1}{2}$ to denote the cube to the right (left) of $x$. Following the discussion of Sec. \ref{css-codes}, our task now is to find certain qubits such that their $z$ values allow the values of $\Phi$ to be determined (that is, the set $S$ described therein). The simplest possible choice is specifying $z_{xyz;1}$ for all vertices. In that case, the value of $\Phi$ can be inverted, since \eqref{eq:z1-by-phi} specifies $\Phi_{x^+y^+z^+}$ in terms of $z_{(xyz;1)}$ and $\Phi$ values on cubes that are closer to the origin than $x^+y^+z^+$, further setting boundary conditions of the form $\Phi_{ijk}=0$ when any of $i,j,k<0$ allows $\Phi$ to be determined uniquely for all cubes. Then, using \eqref{eq:z2-by-phi} the values of all $z$ can be determined based on $\Phi$. This gives all of the information required to obtain the entangling gates which give the ground state of Haah's code. 

\begin{figure}
    \centering
    \begin{tikzpicture}
        \node at (0,0) {\includegraphics[scale=.8]{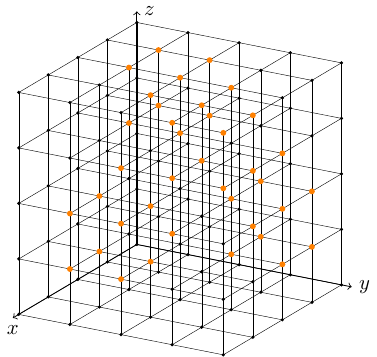}};
        \node at (-2.5,2.5) {(a)};
    \end{tikzpicture}
    \begin{tikzpicture}
        \node at (0,0) {\includegraphics[scale=.65]{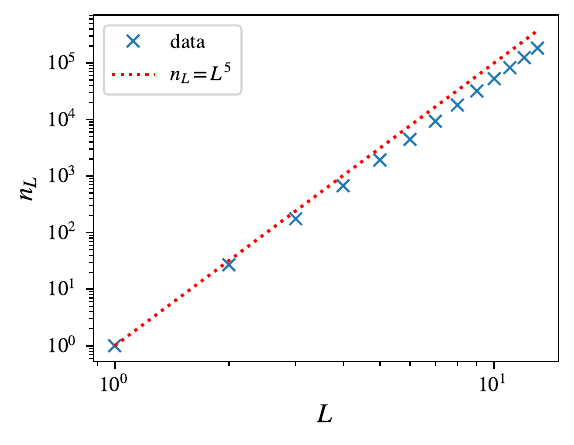}};
        \node at (-3,2.5) {(b)};
    \end{tikzpicture}
    \caption{(a) the $(xyz;2)$ qubits controlling the qubit $(5,5,5;1)$ for the protocol obtaining Haah's code. The controlling qubits are marked in orange. A fractal pattern emerges due to the fractal nature of the logical operators. (b) The total number of two-qubit gates required to obtain Haah's code using our protocol, as obtained numerically. The number is consistent with a total of $O(L^5)$ gates.}
    \label{fig:haahs-code}
\end{figure}

In Fig. \ref{fig:haahs-code} we sketch the form of the controlling qubits which control one of the qubits initialized at $\ket{0}$. We further estimate the number of two-qubit gates required to obtain Haah's code on a cube of length $L$ using this procedure. We obtain results consistent with $O(L^5)$ gates (possibly with logarithmic corrections). We conjecture that this number can be brought down further by a better choice of the qubits initialized at $\ket{+}$.

\section{Outlook}
\label{outlook}
In this work, we presented several protocols for creating topologically ordered states using finite-depth circuits employing simultaneous commuting gates. Our constructions included abelian and certain non-abelian states in 2+1d, as well as higher dimensional CSS-code states. Below we pose several questions for future work.

Perhaps the most significant omission of the present work is the effect of noise on the resulting state. The simplest noise model to consider is random unitary one-qubit gates applied before the all-to-all layers. In the thermodynamic limit, independent random noise applied on each qubit with constant probability $p>0$ is expected to destroy the resulting state (in the sense that the resulting fidelity will vanish). On the other hand, it might be possible to recover the intended ground state by measuring the stabilizers after the state is created and applying an error-correcting circuit. In the problem of random noise applied to the toric code, it is known that for $p$ small enough (but independent of the system size), the state could be recovered \cite{dennis2002topological}. Will it be the case for the protocols discussed here when the preparation is affected by noise?


Finally, interesting questions remain with regard to the construction of more general CSS codes presented in Secs. \ref{css-codes} and \ref{fractons}. Besides the possible experimental realization, it is interesting to ask if there is a general relation between the minimal number of gates required to obtain the code and its code properties (e.g. the rate and distance). Does it have any implication on the realization of good quantum codes \cite{breuckmann2021quantum} using our method?
\section*{Acknowledgements}
We thank Yotam Shapira, Roee Ozery, Ori Grossman, and Gilad Kishony for useful discussions. EB and AS thank the Kavli Institute of Theoretical Physics for their hospitality. This work was supported by grants from the ERC under the European Union’s Horizon 2020 research and innovation programme (Grant Agreements LEGOTOP No. 788715), the DFG (CRC/Transregio 183, EI 519/71), and the ISF Quantum Science and Technology program.
\appendix
\section{Anyons and fusion rules in the gauged string-net model}
\label{app:gauged theory}
Here we review the anyon theory $\C^G$ obtained by gauging an anyonic symmetry $G$ acting on an anyon theory $\C$. Our discussion here is contained entirely in previous work \cite{barkeshli_symmetry_2019,teo_theory_2015}, and simplifications are made for clarity. The process in its full generality is described in \cite{barkeshli_symmetry_2019}. Below we assume that $G$ is abelian and that no symmetry fictionalization occurs in the action of $G$ on $\C$.

The first component for the gauging procedure is the ``defect fusion category" of the anyon theory $\C$. This describes the possible anyons and $G$-defects (those can be seen as vortices such that an anyon $a$ becomes $g(a)$ when it goes around a vortex). For each $g$ there might be more than one defect corresponding to that $g$ action. Physically, this is because an anyon can be attached to the defect. The defect theory $\C^\times_G$ can then be decomposed to theories of anyons which attach to defects as
\begin{equation}
    \C^\times_G=\sum_{g\in G}\C_g
\end{equation}
where $\C_0=\C$ (the anyons with no $G$-defect attached). 

Gauging the symmetry means letting the defects become mobile excitations of the theory. In that case, an anyon $a\in \C$ can only be defined up to actions of the $G$ symmetry: two anyons $a,b$ for which $a=g(b)$ have to be identified. The resulting anyons in the theory $\C^G$ should be thought of as a dyonic excitation comprised of (1) an equivalence class $[a]$ of anyons+defects of $C_G^\times$ that are identified under the action of $G$, and (2) a representation $\pi_a$ of the subgroup $G_a=\qty{g\in G\,|\,g(a)=a}$. One should think of equivalence classes and representations as the magnetic and electric charges, respectively.

To specify the fusion rules, we need to understand how both magnetic and electric charges fuse. For magnetic charges, the rule is that to obtain a equivalence class $[c]$ by fusion $[a]\times[b]$ we must have $\tilde{a}\in[a],\tilde{b}\in[b],\tilde{c}\in[c]$ such that $N_{\tilde{a}\tilde{b}}^{\tilde{c}}=1$. 
As for electric charges, one should note that, in general, the fusion space $V_{ab}^c$ can carry a non-trivial representation $\pi_{a,b;c}$ of $H_{a,b;c}=G_a\cap G_b\cap G_c$ (this can be thought of as the transformation rule of the state having anyons $a,b,c$ fusing together to the vacuum). The possible representations attached to the element $\tilde{c}$ should be obtained as the tensor product 
\begin{equation}
\pi_{\tilde{c}}|_{H_{\tilde{a}\tilde{b};\tilde{c}}}=\pi_{\tilde{a}}|_{H_{\tilde{a}\tilde{b};\tilde{c}}} \otimes \pi_{\tilde{b}}|_{H_{\tilde{a}\tilde{b};\tilde{c}}} \otimes \pi_{\tilde{a},\tilde{b};\tilde{c}}
\end{equation}
where $|_{H_{\tilde{a}\tilde{b};\tilde{c}}}$ is the restriction of the representation on ${H_{\tilde{a}\tilde{b};\tilde{c}}}$.

\subsubsection*{Example: gauging the electric-magnetic duality of the toric code}
We consider the example in which the theory $\C={1,e,m,\psi}$ is the toric code, and the symmetry action $g$ takes $e\leftrightarrow m$. We have two types of defects $\C_g={\sigma,\sigma_e}$ where $\sigma$ is the ``bare'' defect that interchanges $e$ and $m$, and $\sigma_e$ is $\sigma\times e$ (or equivalently $\sigma\times m$). The fusion rules of the defects are given by
\begin{align}
    \sigma\times\sigma=\sigma_e\times\sigma_e&=1+\psi,&\sigma\times\sigma_e=e+m.
\end{align}
We now gauge the symmetry. Under gauging, the distinction between $e,m$ is no longer defined, and we consider $[e]$ as a single anyon. All other anyons can carry an additional non-trivial symmetry charge $z$ (corresponding to the $\{1,-1\}$ representation of $G=\ZZ_2$). The anyons of the theory are then
\begin{equation}
\begin{aligned}
    &1,z,\psi,\psi z,\sigma,\sigma z, \sigma_e,\sigma_e z,[e].
\end{aligned}
\end{equation}
The fusion rules of the abelian anyons in the theory are determined by $\psi\times\psi=1,z\times z=1$. The other fusion rules are given by
\begin{equation}
\begin{aligned}
    [e]\times[e] &= 1+z+\psi+\psi z,\\
    \sigma\times\sigma_e&=[e], \\
    [e]z&=[e], \\
    \sigma\times\sigma &= 1+\psi, \\
    \sigma_e\times \sigma_e &= 1+\psi z
\end{aligned}
\end{equation}
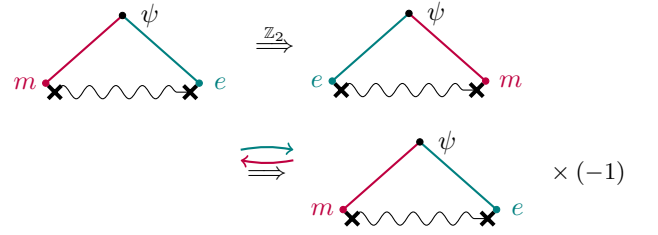
\begin{figure}
    \tikzset{cross/.style={cross out, draw=black, line width=1.5, minimum size=2*(#1-\pgflinewidth), inner sep=0pt, outer sep=0pt},
    cross/.default={4pt}}
    \centering
    \begin{equation*}
    \begin{aligned}
    \vcenter{\hbox{
    \begin{tikzpicture}[scale=.6,every node/.append style={font={\normalsize}}]
    \draw[decorate,decoration=snake] (-1.5,0)node[cross] {} -- (1.5,0)node[cross] {};
    \draw[thick,teal] (1.7,.2)node[circle,right] {$e$} -- (0,1.7)node[black,right,circle] {$\psi$};
    \draw[thick,purple] (-1.7,.2)node[left] {$m$} -- (0,1.7);
    \filldraw[teal] (1.7,.2) circle (.07);
    \filldraw[purple] (-1.7,.2) circle (.07);
    \filldraw[black] (0,1.7) circle (.07);
    \end{tikzpicture}
    }} \overset{\ZZ_2}{\Longrightarrow}
    &\vcenter{\hbox{
    \begin{tikzpicture}[scale=.6,every node/.append style={font={\normalsize}}]
    \draw[decorate,decoration=snake] (-1.5,0)node[cross] {} -- (1.5,0)node[cross] {};
    \draw[thick,purple] (1.7,.2)node[circle,right] {$m$} -- (0,1.7)node[black,right,circle] {$\psi$};
    \draw[thick,teal] (-1.7,.2)node[left] {$e$} -- (0,1.7);
    \filldraw[purple] (1.7,.2) circle (.07);
    \filldraw[teal] (-1.7,.2) circle (.07);
    \filldraw[black] (0,1.7) circle (.07);
    \end{tikzpicture} 
    }}\\
    \overset{
    \begin{tikzpicture}[scale=.7]
        \draw[->,thick,teal] (-.5,.1) to[out=10,in=170] (.5,.1);
        \draw[->,thick,purple] (.5,-.1) to[out=-170,in=-10] (-.5,-.1);
    \end{tikzpicture}
    }{\Longrightarrow}
    &\vcenter{\hbox{
    \begin{tikzpicture}[scale=.6,every node/.append style={font={\normalsize}}]
    \draw[decorate,decoration=snake] (-1.5,0)node[cross] {} -- (1.5,0)node[cross] {};
    \draw[thick,teal] (1.7,.2)node[circle,right] {$e$} -- (0,1.7)node[black,right,circle] {$\psi$};
    \draw[thick,purple] (-1.7,.2)node[left] {$m$} -- (0,1.7);
    \filldraw[teal] (1.7,.2) circle (.07);
    \filldraw[purple] (-1.7,.2) circle (.07);
    \filldraw[black] (0,1.7) circle (.07);
    \end{tikzpicture}
    }}\times (-1)
    \end{aligned}
    \end{equation*}
    \caption{Fusion of $\sigma_e\times\sigma_e$ to $\psi$ transforms non-trivially under the $G$ action}
    \label{fig:sigma-psi-fusion}
\end{figure}
The last line merits additional attention; the additional charge attached to $\psi$ is a result of the fusion space transforming non-trivially: when a defect with $e$ near it and a defect with $m$ near it fuse to obtain $\psi$, the action of $G$ exchanges $e\leftrightarrow m$, and results in an additional $-1$ phase (see Fig. \ref{fig:sigma-psi-fusion}). We also note that the resulting fusion rules match those of two copies of an Ising anyon theory, allowing us to identify the resulting anyon theory rules as ${\rm Ising}\times \overline{\rm Ising}$ \cite{teo_theory_2015}.
\section{Convergence of the algorithm for a CSS code}
\label{app:algorithm}
In Sec. \ref{css-codes} we described an algorithm for obtaining the set $S$ satisfying \eqref{eq:s-dim} and \eqref{eq:s-dim-2}. To guarantee that the algorithm converges, we need to show that for a set $S$ such that $\abs{S}<\im \AA$ and \eqref{eq:s-dim} is satisfied, there always exists some qubit $r$ such that \eqref{eq:s-dim} is satisfied for $S\cup\qty{r}$. We now complete the analysis by proving the above statement.

Assume that for some $S$ satisfying \eqref{eq:s-dim}, for any $r\notin S$ the set $S_r=S\cup \qty{r}$ satisfies
\begin{equation}
    \abs{\im{\pi_{S_r}\AA}}=\abs{S_r}-1.
\end{equation}
This means that there exists a non-zero vector $\vec{v_r}\in \ft^{\abs{S_r}}$, such that for any $\vec{z}\in \im{\pi_{S_r}\AA}$ we have $\vec{v_r}\cdot \vec{z}=0$. The vector $\vec{v_r}$ cannot be supported only on $S$, because then we have $\abs{\im{\pi_{S}\AA}}<\abs{S}$. The values of $v_r$ allow us to determine the value of the qubit $r$ given the qubits in $S$. Explicitly, we can define an operator $\MM\in\ft^{N\times \abs{S}}$ by
\begin{equation}
    \MM_{i,j} =\begin{cases}
        (\pi_S)_{j,i}& j\in S \\
        (\vec{v_j})_i& j\notin S
    \end{cases},
\end{equation}
with $(\vec{v_j})_i$ being the $i$'th index of $\vec{v_j}$. Importantly, since $v_r$ determines the value of the qubit $r$ given the qubits in $S$ for any $\vec{z}\in\im{\AA}$, we have $\im{\MM}=\im{\AA}$. In particular, this means that $S$ satisfies $\abs{S}=\abs{\im \AA}$ as required.

\bibliography{bibliography.bib}
\end{document}